\documentclass[9pt,twocolumn,twoside,lineno]{pnas-new}
\usepackage[utf8]{inputenc}
\usepackage{graphicx}
\usepackage{amsmath}
\usepackage{amsfonts}
\usepackage{amssymb}
\DeclareMathOperator{\erf}{erf}

\usepackage{float}
\usepackage{placeins}
\usepackage{graphicx}
\usepackage{wrapfig}
\usepackage[font={small}]{caption}
\usepackage{subfig}
\usepackage{threeparttable, tablefootnote}
\usepackage{adjustbox}
\usepackage{siunitx}

\templatetype{pnasresearcharticle} % Choose template 
\title{Cerebral cortical communication  overshadows computational energy-use, but these combine to predict synapse number}

\author[a,1]{William B Levy}
\author[b]{Victoria G. Calvert} 

\affil[a]{Department of Neurosurgery and Department of Psychology; University of Virginia; Charlottesville, VA 22903; USA}
\affil[b]{College of Arts and Sciences; University of Virginia; Charlottesville, VA 22903; USA}

\leadauthor{Levy} 

\significancestatement{Engineers hold up the human brain as a low energy form of computation. However from the simplest physical viewpoint, a neuron’s computation cost is remarkably larger than the best possible bits/J -- off by a factor of $10^8$. Here we explicate, in the context of energy consumption, a definition of neural computation that is optimal given explicit constraints. The plausibility of this definition as Nature’s perspective is supported by an energy-audit of the human brain. The audit itself requires certain novel perspectives and calculations revealing that communication costs are 35-fold computational costs.}

\authorcontributions{WBL conceptualized the study, developed the theoretical aspects and their description. WBL and VC developed the energy-audit and its description.}
\authordeclaration{There are no competing interests}
\correspondingauthor{\textsuperscript{1} E-mail: wbl@virginia.edu}


\begin{abstract}

Darwinian evolution tends to produce energy-efficient outcomes. On the other hand, energy limits computation, be it neural and probabilistic or digital and logical.  Taking a particular energy-efficient viewpoint, we define neural computation and make use of an energy-constrained, computational function. This function can be optimized over a variable that is proportional to the number of synapses per neuron. This function also implies a specific distinction between ATP-consuming processes, especially computation \textit{per se} vs the communication processes including action potentials and transmitter release. Thus to apply this mathematical function requires an energy audit with a partitioning of energy consumption that differs from earlier work. The audit points out that, rather than the oft-quoted 20 watts of glucose available to the brain \cite{sokoloff1960metabolism,sawada2013synapse}, the fraction partitioned to cortical computation is only 0.1 watts of ATP. On the other hand at 3.5 watts, long-distance communication costs are 35-fold greater. Other novel quantifications include (i) a finding that the biological vs ideal values of neural computational efficiency differ by a factor of $10^8$ and (ii) two predictions of $N$, the number of synaptic transmissions needed to fire a neuron (2500 vs 2000).  

\end{abstract}

\dates{This manuscript was compiled on \today}
\doi{\url{www.pnas.org/cgi/doi/10.1073/pnas.XXXXXXXXXX}}

\begin{document}

\maketitle
\thispagestyle{firststyle}
\ifthenelse{\boolean{shortarticle}}{\ifthenelse{\boolean{singlecolumn}}{\abscontentformatted}{\abscontent}}{}

% If your first paragraph (i.e. with the \dropcap) contains a list environment (quote, quotation, theorem, definition, enumerate, itemize...), the line after the list may have some extra indentation. If this is the case, add \parshape=0 to the end of the list environment.

%..........!!! SEPT 28 NEW INTRO........
\dropcap{T}he purpose of the brain is to process information, but that leaves us with the problem of finding appropriate definitions of information processing. We assume that given enough time and given a  sufficiently stable environment  (e.g., the common internals of the mammalian brain), then Nature’s constructions approach an optimum. The problem is to find which function or combined set of functions are optimal when we incorporate empirical values into the function(s). The initial example in neuroscience is \cite{levybaxter96}, which shows that information capacity is far from optimized, especially in comparison to the optimal information per joule which is in much closer agreement with empirical values. Whenever we find such an agreement between theory and experiment, we conclude that this optimization, or near optimization, is Nature’s perspective. Using this strategy, we and others seek quantified relationships with particular forms of information processing and require that these relationships are approximately optimal \cite{levybaxter96,alexander1996optima,vijayberry01, levy2002energy,sterling2015principles, balasubramanian2015heterogeneity,stone2018principles}. A recent theoretical development identifies a candidate optimal computation at the level of  single neurons \cite{levybergersungkar}. To apply this theory requires understanding certain neuronal energy expenditures. Here the focus is on the energy budget of the human cerebral cortex and its primary neurons. The energy audit here differs from the premier earlier work \cite{attwell2001energy} in two ways: the brain considered here is human not rodent, and the audit here uses a partitioning  motivated by the information-efficiency calculations rather than the classical partitions of cell biology and neuroscience \cite{attwell2001energy}. Importantly, our audit reveals greater joule-use by communication than by computation. This observation in turn generates new insights into the optimal synapse number. Specifically, the bits/J optimized computation must provide a sufficient bits/sec to the axon and presynaptic mechanism to justify the great expense of timely communication. Simply put from the optimization perspective, we assume evolution does not build a costly communication system and then fails to supply it with an appropriate bits/sec to justify its costs. The bits/J are optimized over $N$ when $N\approx 2000$, the number of synaptic activations per interpulse interval (IPI) for one neuron, where \textit{a priori}  $N$ is assumed to equal the number of synapses per neuron times the success rate of synaptic transmission (an estimated $2500$).  

To measure computation, and to partition out its cost, requires a suitable definition at the single neuron level. Rather than the generic definition ‘any signal transformation’ \cite{vijayberry01}, or the neural-like ‘converting a multivariate signal to scalar signal’, we conjecture a more detailed definition \cite{levybergersungkar}. To move towards this definition,  note two important brain functions: estimating what is present in the sensed world and predicting what will be present, including what will occur as the brain commands manipulations. Then, assume that such macroscopic inferences arise by combining single neuron inferences. Such a neuron performs the same type of inference as the macroscopic problem; i.e., conjecture a neuron performing microscopic estimation (or prediction). Instead of sensing the world, a neuron's sensing is merely its capacitive charging due to recently active synapses. Using this sampling of  total accumulated charge over a particular elapsed time, a neuron implicitly estimates the value of its local latent variable, a variable defined by evolution and developmental construction \cite{levybergersungkar}. Applying an optimization perspective, which includes implicit Bayesian inference, a sufficient statistic, maximum-likelihood unbiasedness, as well as energy costs, \cite{levybergersungkar} produces a quantified theory of single neuron computation. A result of this theory is the definition of the optimal IPI probability distribution. Motivating IPI coding is this fact: The use of constant amplitude signaling, e.g. action potentials, implies that all information can only be in IPIs. Therefore, no code can outperform an IPI code, and it can only equal an IPI code in bit-rate if it is one-to-one with an IPI code. In neuroscience, an equivalent to IPI codes is the instantaneous rate code where each  message is $IPI^{-1}$. In communication theory, a discrete form of IPI coding is called  differential pulse position modulation \cite{mayer1959feasibility}; \cite{bergerlevy10} explicitly introduced a continuous form of this coding as a neuron communication hypothesis, and it receives further development in \cite{sungkar2017capacity}. 

The Results recall and further develop  earlier work concerning a certain optimization that defines IPI probabilities \cite{levybergersungkar}. An energy audit is required to use these novel developments. Combining the theory with the audit leads to two outcomes: (i) the optimizing $N$ serves as a consistency check on the audit and (ii) future energy audits for individual cell types will predict $N$ for that cell type, a test of the theory. Specialized approximations here that are not present in earlier work \cite{attwell2001energy} include the assumptions that (i) all neurons of cortex are pyramidal neurons, (ii) pyramidal neurons are the inputs to pyramidal neurons, (iii) a neuron is under constant synaptic bombardment, and (iv) a neuron's capacitance must be charged 16 mV from reset potential to threshold to fire. 

\begin{figure*}[htpb]

\adjincludegraphics[width=20cm,Clip={.00\width} {.17\height} {0.15\width} {.01\height}]{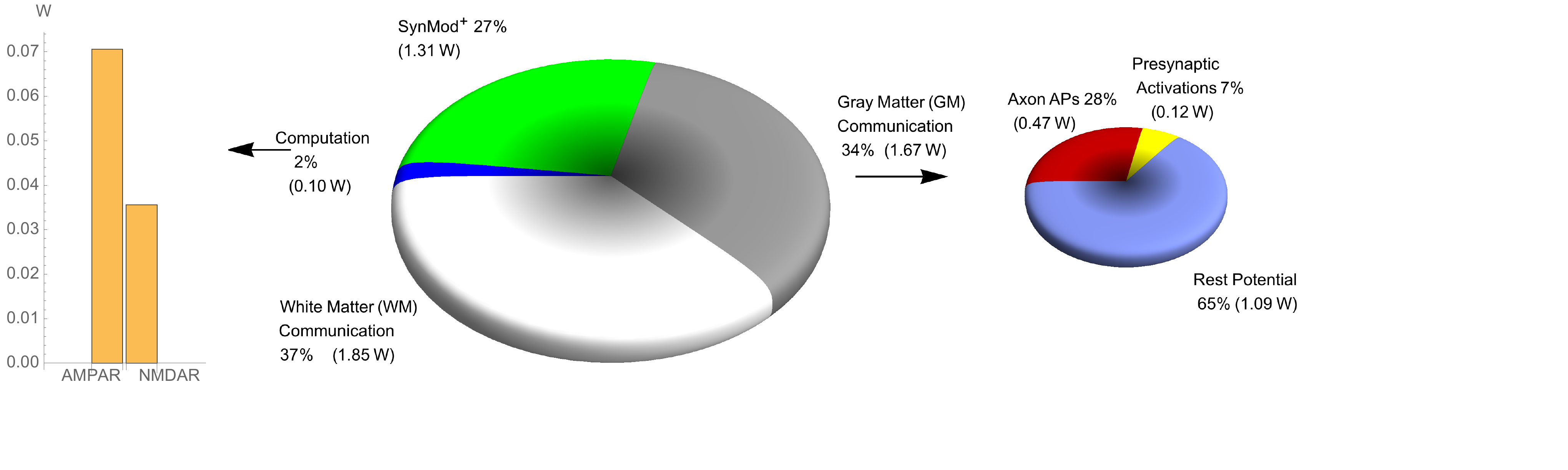}

 \caption{Computation costs little compared to communication. Communication alone accounts for more ca. two thirds of the available 4.94 ATP-W (Table 1), with slightly more consumption due to WM than  GM (big pie chart). Computation, the smallest consumer, is subpartitioned by the two ionotropic glutamate receptors (bargraph). $SynMod^+$ includes astrocytic costs, process extension, process growth, axo- and dendro-plasmic transport of the membrane building blocks, and time-independent housekeeping costs (although this last contributor is a very small fraction). The small pie chart sub-partitions GM communication. See Results and Methods for details. WM communication includes its maintenance and myelination costs in addition to resting and action potentials.
}
 
 \label{piediagrams}
\end{figure*}

Following the audit, the reader is given a perspective that may be obvious to some, but it is rarely discussed and seemingly contradicts the engineering literature (but see \cite{balasubramanian2015heterogeneity}). In particular, a neuron is an incredibly \textit{inefficient} computational device in comparison to a physical analog. It is not just a few bits/J away from optimal, but off by a huge amount, a factor of $10^8$. The theory resolves the efficiency issue using a modified optimization perspective. Activity dependent communication and synaptic modification costs force upward optimal computational costs. In turn, the bit-value of the computational energy expenditure  is constrained to a central-limit like result: Every doubling of $N$ can produce no more than 0.5 bits. In addition to (i) explaining the $10^8$ excessive energy-use other results here include (ii) identifying the largest ‘noise’ source limiting computation, which is the signal itself, and  (iii) partitioning the relevant costs, which may help engineers redirect focus towards computation and communication costs rather than the 20 W total brain consumption as their design goal.

\section*{Results}

\subsection*{Energy audit}

\subsubsection*{ATP use for computation and communication}
Microscopic energy costs are based on bottom up calculations for ATP consumption \cite{attwell2001energy}.  36,000 J/molATP, \cite{nath2016thermodynamic} implies watts. As derived below, computation consumes ca. 0.1 ATP-watts/cortex or  \textit{one two-hundredth of the nominal and oft quoted 20 watts that would be produced by complete oxidation of the glucose taken up by the brain} \cite{sokoloff1960metabolism}. Fig \ref{piediagrams} compares cortical communication costs to computational costs. Also appearing is an energy consumption labeled ($SynMod^+$). What \cite{howarth2012updated} calls 'housekeeping' is a hypothesis on its part; an alternative hypothesis  is inspired by and consistent with results from developing brain \cite{engl2017non}. This category seems to be dominated by costs consistent with synaptogenesis (e.g.,growth and  process extension via actin polymerization and via new membrane incorporation, membrane synthesis and its axo- and dendro-plasmic transport, and astrocytic costs); a small fraction of  $SynMod^+$ is time-dependent 'maintenance'. Here $SynMod^+$ is calculated by subtracting the bottom-up calculated communication and computation ATP consumption from available ATP, a top-down empirical partitioning (Table \ref{mainwatts}). 

%What follows is discussion  move or remove ??
For some, the rather large cost of communication might be surprising but apparently is necessary for suitable signal velocities and information rates \cite{levybaxter96,crotty2006metabolic,sangrey2004analysis,sterlingvijay,perge2012}. Combining gray matter (GM) communication costs with the total white matter (WM) costs accounts for 71\%, 3.52 W (Fig \ref{firingrateandenergy}), of the total 4.94 ATP-watts/cortex, compared to 2\% for computation. Supposing that all WM costs are essentially communication costs (including oligodendrocyte/myelination costs), then the ratio for communication vs computation is 35:1.

\begin{table}[htpb]
\centering
\caption{Rudimentary partitioning, glucose to ATP}
\label{mainwatts}
%\begin{threeparttable}

 \centering %
\begin{tabular}{lllll}
\toprule
Brain/Region & Watts & Unoxidized& Heat  & ATP- \\
(weight)&(complete  & (equivalent & watts & watts \\
 & oxidation) & watts)   &   \\\toprule %
whole brain (1495 g)& 17.0 & 1.86 & 8.89 & 6.19 \\
\midrule
\;\;cerebellum (154 g)  & 1.77 & 0.19 & 0.93 & 0.65 \\ 
\;\;other regions\,  (118 g) & 1.65 & 0.18 & 0.87 & 0.60  \\
\midrule
\;\;forebrain cortex (1223 g): &&&& 4.94\\
\;\;\;\;white (590 g)  & 5.07 & 0.56 & 2.66 &1.85 \\
\;\;\;\;gray \, (633 g) & 8.45 & 0.93 & 4.43 &3.09 \\ 
%\midrule
%Bottom-up calculated ATP requirements\\ @ 1.6 Hz firing rate\\
%\midrule

%\;\;\;\;gray: \\
%\quad\; communication & & & & 2.75 \\
%\quad\; computation  & & & & 0.17 \\
%\quad\; $SynMod^+$ ATP demands & & & & 0.17\tnote{$\dagger\dagger$}   \\

%\midrule
\bottomrule 
\end{tabular}
\addtabletext{See Methods and SI Appendix (Tables) for details and citations}
%\begin{tablenotes} \footnotesize
%\item The footnotes and derivations for this table can be found in Table S2.
%\item[$\diamond$]Watts, including the whole brain value, are based on regional glucose-uptake values from \cite{graham2002fdg}, assuming complete oxidation, and $2.8$ MJ/mol\;glucose \cite{nelson2008lehninger} ; regional masses from \cite{azevedo2009equal}. 
%\item[$\diamond\diamond$] Also assuming complete oxidation of this non-oxidized glucose.
%\item[$\star$]Using Nath's torsional mechanism \cite{nath2016thermodynamic}\cite{nath2010beyond}  which incorporates mitochondrial leak
%\item[$\dagger$]$36$  kJ/molATP \cite{nath2016thermodynamic}\cite{nath2009energy}
%\item[$\star\star$]Including basal ganglia, thalamus, brainstem, etc. The missing mass is ventricular. See Table 1 and the accompanying footnotes for more information.
%\item[$\dagger\dagger$]Assuming that $SynMod^+$ consumes the same amount of energy as computation.
%\end{tablenotes}
%\end{threeparttable}
\end{table}
%\FloatBarrier

\subsubsection*{Computation costs in the human brain}
The energy needed to recover ion-gradients from the total excitatory synaptic current-flows/IPI determines the cost of computation for that IPI. Various quantitative assumptions feeding into subsequent calculations are required (see Methods and SI Appendix), but none are more important than the generic assumption that the average firing-rate of each input to a neuron is the same as the average firing-rate out of that neuron \cite{levy2002energy}. Via this assumption, and assuming $10^4$ synapses per neuron and a 75\% failure rate, the aggregate effects of inhibition, capacitance, and postsynaptic K$^+$ conductances are implicitly taken into account. This aggregation is possible since increases of any of these parameters merely lead to smaller depolarizations per synaptic activation but cause little change in synaptic current flow per excitatory synaptic event. Indeed, such attenuating effects are needed to make sense of several other variables. A quick calculation helps illustrate this claim.

An important starting point for computational energy cost is the average number of excitatory synaptic activations to fire a cortical neuron. Assume the neuron is a pyramidal neuron and that its excitatory inputs are other pyramidal neurons. Therefore, the mean firing rate of this neuron is equal to the mean firing rate of each input. Thus, threshold will be the number of  input synapses times quantal success rate \cite{levy2002energy}; i.e., ca. 
$10^4 \cdot   0.25=2500=N$ because on average each input fires once per IPI out. Even after accounting for quantal synaptic failures, inhibition is required for consistency with 2500 excitatory events propelling the 16 mV journey from reset to threshold. Activation of AMPARs and NMDARs provides an influx of three Na$^+$'s for every two K$^+$ that flow out. With an average total AMPAR conductance of 200 pS, there are 114.5 pS of Na$^+$ per synaptic activation (SA). Multiplying this conductance by the 110 mV driving force on Na$^+$ and by the 1.2 msec SA duration  yields $15.1$ fC per SA. Dividing this total Na$^+$ influx by 3 compensates for the 2 K$^+$ that flow out for every 3 Na$^+$ that enter; thus, the  net charge influx is 5.04 fC/SA. We assume that the voltage-activated,  glutamate-primed NMDARs increases this net flux by a factor of 1.5, yielding 7.56 fC/SA (see Methods and SI Appendix, Tables S3, S4, and S5 for details and ATP costs). Taking into account the 2500 synaptic activations per IPI yields 18.9 pC/IPI. Using a 750 pF value for a neuron's capacitance, this amount of charge would depolarize the membrane potential 25.2 mV rather than the desired 16 mV. Thus, the excitatory charge influx must be opposed by inhibition and K$^+$ conductances to offset the total 7.56 fC net positive influx. Most simply, just assume a divisive inhibitory factor of 1.5. Then the numbers are all consistent, and the average depolarization is 6.4 $\mu$V per synaptic activation. Because each net, accumulated charge requires one ATP to return the three Na$^+$'s and 2 K$^+$'s, the computational cost of the 16 mV depolarization is $6.67 \cdot 10^{-12}$ J/neuron/spike. I. e., required computational power for each neuron spike of  cortex  $6.67 \cdot 10^{-12}\cdot 1.5 \cdot 10^{10}=0.10$ W.

\subsubsection*{Communication costs}
As quantified in Methods (see also SI Appendix, Tables S3 and S5), the GM long-distance communication cost of $1.67$ W (Fig 1) includes the partitioned costs of axonal resting potential, APs, and presynaptic transmission (neurotransmitter recycling and packaging, vesicle recycling, and calcium extrusion). The neurotransmission costs assume a 1 Hz mean neuron firing rate and a $75\%$ failure rate. Next using \cite{howarth2012updated},  the calculation assumes one vesicle is released per non-failed AP. Differing from \cite{howarth2012updated} while closer to earlier work \cite{attwell2001energy}, assume there is the same Ca-influx with every AP \cite{stevens1995facilitation}. Further,  also use a more recent measurement of Na$^+$-K$^+$ overlapping current flows of the AP, 2.38 \cite{hallermann2012state}. Of all the difficult but influential estimates, none is more challenging and important than axonal surface area; see Methods.

\subsubsection*{Firing Rate}
In regard to average firing rate, we postulate an average value of one pulse per neuron per decision-making interval (DMI), which we assume as 1 sec.

\FloatBarrier
 \begin{figure}[htbp]
\centering
\adjincludegraphics[width=8.6cm,Clip={.00\width} {0.007\height} {0.0\width} {.00\height}]{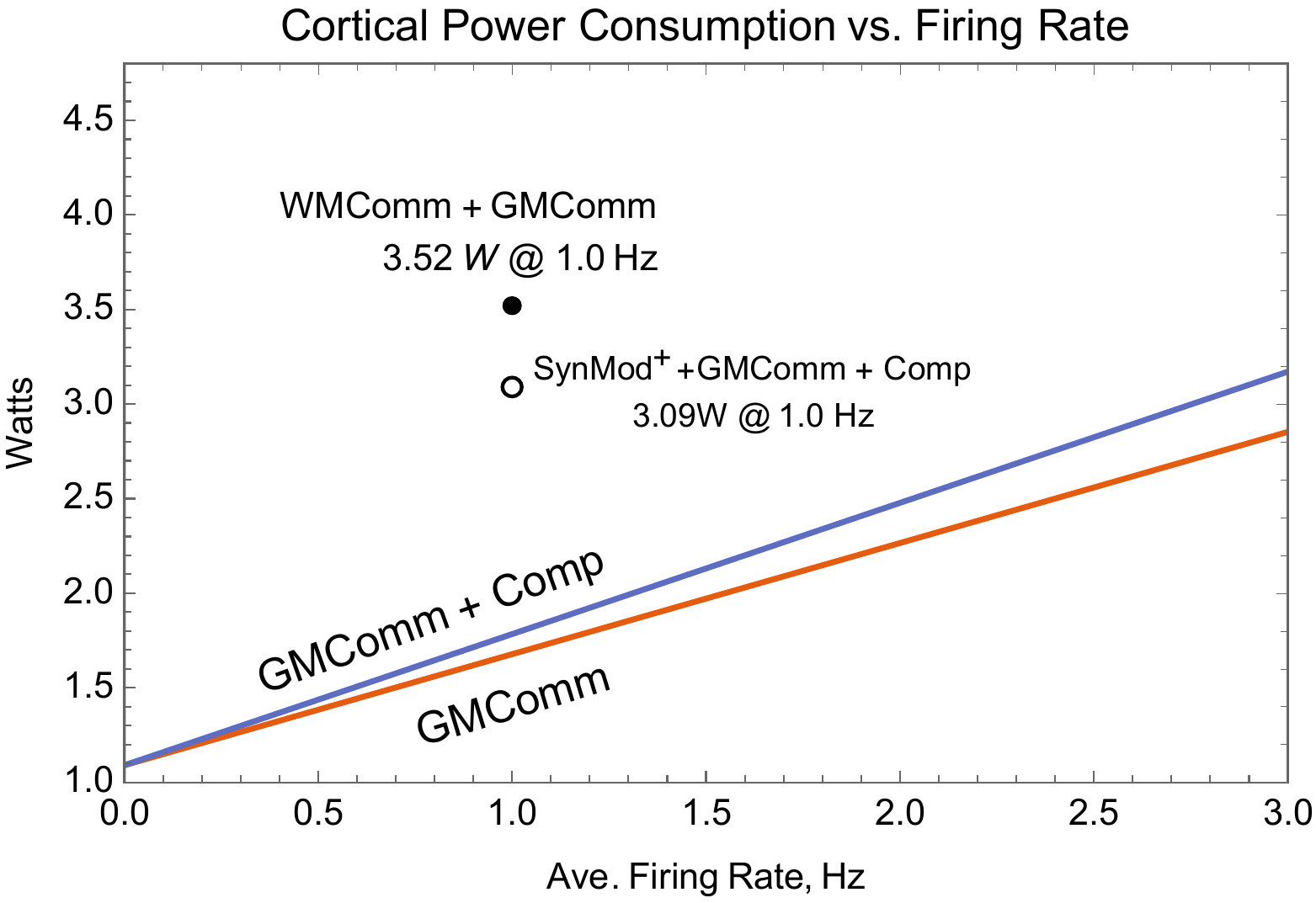}
% \includegraphics[scale=0.535]{firing_rate_fig2_oct19_2020.pdf}
% \end{figure}
%\FloatBarrier
%\begin{figure}[htbp] 
\caption{Energy-use increases linearly with average firing-rate, but for reasonable rates, computation (Comp) costs much less than communication (Comm). Comparing the bottom (red) curve (GM communication costs) to the top (blue) curve (GM communication cost plus computational costs), illustrates how little computational costs increase relative to communication costs. The y-intercept value is 1.09 W for resting potential. The unfilled circle plotting $SynMod^+$+GMComm+Comp adds the 1.32 W of GM $SynMod^+$ to the 1.77 W of GMComm+Comp $@$ 1 Hz. The filled circle, labeled WMComm + GMComm, shows the value of the combined communication cost, cortical GM at 1 Hz, and the total cortical white matter cost. See Methods for further details.}
%Energy-use increases linearly with average firing-rate, but for reasonable rates, computation (Comp) costs much less than communication (Comm). Comparing the bottom (blue) curve (GM communication costs) to the top (red) curve (GM communication cost plus computational costs), illustrates how little computational costs increase relative to communication costs. The large y-intercept value is 1.8 W for resting potential plus 0.1 W for a constant consumption by $SynMod^+$. The small point labeled GMAPOther+GMComm+Comp adds 0.07 W of AP-dependent $SynMod^+$ to the GMComm+Comp curve, 0.17 W + 2.75 W $@$ 1.6 Hz. The large point, labeled WM(Comm + Other) + GMComm, shows the value of the combined communication cost, cortical GM at 1.6 Hz plus the total cortical white matter (WM) cost. See Table \ref{mainwatts}, and Methods for further details.}
\label{firingrateandenergy}
\end{figure}
\FloatBarrier

%In contrast, \cite{lennie2003cost} prefers a human average firing rate closer to $0.1$ Hz than $1$ Hz because his calculations do not have enough energy available. However, the calculations here provide a comfortable margin for the spike-independent energy uses. Even a rate of 1.6 Hz might be considered (see Table S2 in SI).
As Fig \ref{firingrateandenergy} indicates, the combined WM and GM communication cost at 1 Hz is 3.52 W. Computational costs are only a very small fraction of frequency-dependent costs. Calculation of $SynMod^+$ is not possible and, as explicated in Discussion, we discredit the ouabain manipulation others \cite{attwell2001energy,lennie2003cost} use to estimate it. The value here  is arrived at by differencing the calculated and measured costs from the available energy (see SI Appendix, Fig S1).

 Using a firing-rate of $1$ Hz and $1.5\cdot 10^{10}$ neurons/cortex, a bottom up calculation for the excitatory postsynaptic ion-flux per AP per cortex yields 0.10 W. The linear relationship between firing rate and energy consumption has a substantial baseline energy consumption of 1.09 W (y-axis intercept).  Apparently resting axon conductance \cite{raastad2019slow} is required for a resting potential and stable behavior \cite{faisal2007stochastic}. 
In the case of the dendrite, computational costs are zero at zero firing rate, a theoretical limit result which, as argued earlier, is a nonsense practical situation. However, dendritic leak is assumed to be essentially zero since we assume, perhaps controversially (cf. \cite{attwell2001energy}), that a cortical neuron is under constant synaptic bombardment and that all dendrosomatic conductances are due to synaptic activation and voltage-activated channels. That is, a neuron resets after it fires and immediately starts depolarizing until hitting threshold. 

Computational costs are very sensitive to failure rates, which for this figure are fixed  at 75\%, whereas communication is only slightly sensitive to the synaptic failure rate (see below for more details).

\subsubsection*{An energy-use partitioning based on glucose oxidation}
The oft-repeated brain energy consumption of 20 W is not simply the cost of computation and communication, thus requiring an appropriate partitioning (see Table \ref{mainwatts}). The 17 W of glucose potential energy from recent PET scan research \cite{graham2002fdg} replaces Sokoloff's 20 W from the 1950s. The PET scan research produces regional per-gram values, and these values are scaled by the regional masses \cite{Herculano_Houzel_2011}, allowing regional estimates, Table S1. Arguably, 11\% of the total glucose uptake is not oxidized \cite{fox1988nonoxidative}, (some arteriovenous differences obtain a smaller percentage; see SI Appendix). After removing the 8.89 W for heating, there are only 6.19 ATP-W available to the whole brain. The regional partitioning implies cerebral gray consumes 3.09 ATP-W, split between computation, communication, and $SynMod^+$.  After direct calculation of communication and computational costs, the remaining GM energy is allocated to $SynMod^+$.

\subsubsection*{A specialized partitioning}
 The ultimate calculation of bits/J requires a bipartite partition of action potential-driven costs: Those independent of $N$, {\large $A:= \mathcal{E}_{\texttt{WMAP}}+\mathcal{E}_{\texttt{GMAxAP}}+\mathcal{E}_{\texttt{SynModGrow}}\approx 1.23+0.45+ 1.08 = 2.76  $}
J/sec/cortex vs those proportional to $N$,
{\large $B:= \mathcal{E}_{\texttt{COMP}}+\mathcal{E}_{\texttt{Pre}}+\mathcal{E}_{\texttt{NSynMod}}+\mathcal{E}_{\texttt{PreNaAP}}
\approx 0.10+ 0.11 + 0.11 + 0.02= 0.34$} J/sec/cortex. 
For the three non-$N$ contributers, WMAP is white matter action potential-dependent; GMAxAP is gray matter action potential-dependent; $SynModGrow$ are all the functions that underlie synaptogenesis and shedding as well as maintenance as driven by firing. For the four $N$-proportional functions, Comp is postsynaptic ionotropic activation;  Pre is presynaptic Ca$^{2+}$ and transmitter release functions; PreNaAP is partial presynaptic depolarization driven by the axonal AP; and NSynMod is $N$-proportional synaptic modifications including synaptic metabotropic activation. Note that 
 $ \mathcal{E}_{\texttt{SynMod}^{+}}=\mathcal{E}_{\texttt{SynModGrow}}+\mathcal{E}_{\texttt{NSynMod}}+\mathcal{E}_{\texttt{Plus}}\approx
 1.08 +0.11 +0.12$, where $\mathcal{E}_{\texttt{Plus}}$ is the time-dependent maintenance cost. Finally, to use $A$ and $B$, they are rescaled to J/IPI/neuron (divide by number of neurons firing in one sec) and, additionally for $B$, a rescaling to dependence on synapse number (multiply by $N\div 2500$, thus, 
 $\mathcal{E}(\Lambda,T):=(A+N\cdot B/2500)\cdot E[T]\div n$ where $E[T]$ is the average IPI and $n$ is the number of cortical neurons.

\subsection*{A baseline for maximally efficient computation}

\subsubsection*{A simplistic model relates physics to neuroscience}
For the sake of creating a baseline, initial comparison and for further understanding of just what "computation" can mean, suppose a neuron's computation is just its transformation of inputs to outputs. Then, quantifying the information passed through this transformation (bits/sec) and dividing this information rate by the power (W = J/sec) yields bits/J. This ratio is our efficiency measure. In neuroscience, it is generally agreed that Shannon's mutual information ($MI$) is applicable for measuring bit-rate of neural information processing, neural transformations, or neural communication, e.g., \cite{ bialek91,laughlin98, abshire, dayanabbott, vijayberry01, levy2002energy, niven07,harrisattwell15}. Specifically, using mutual information and an associated rate of excitatory postsynaptic currents of a single neuron produces a comparison with the optimal bits/J for computation as developed through physical principles. To understand the analogy with statistical mechanics, assume the only noise is wideband thermal noise, k$\mathcal{T}$ $\approx 4.3 \cdot 10^{-21}$ J (Boltzmann's constant times absolute  temperature, $\mathcal{T}=310$ K). The bits/J ratio can be optimized to find the best possible energetic cost of information, which is $(k\mathcal{T}\ln2)^{-1}$, the Landauer limit \cite{landauer61}.

\begin{figure}[tph]
    \centering
   \adjincludegraphics[width=8.6cm,Clip={.00\width} {0.18007\height} {0.0\width} {.2\height}]{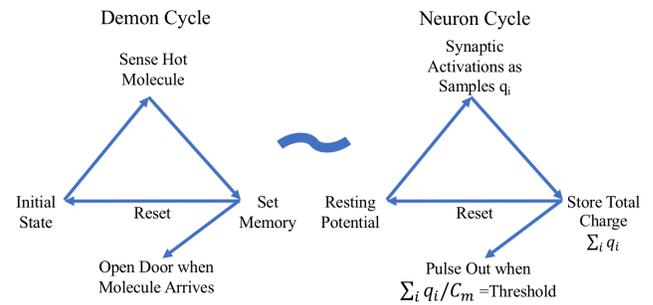}

    \caption{Maxwell's demon cycle is analogous to the neuron's computational cycle.  The initial state in the demon cycle is equivalent to the neuron at rest. The demon sensing fast molecules is analogous to the synaptic activations received by the neuron. Whereas the demon uses energy to set the memory and then opens the door for a molecule, the neuron stores charge on the membrane capacitance (C$_{\textrm{m}}$) and then pulses out once this voltage reaches threshold. Simultaneous with such outputs, both cycles then reset to their initial states and begin again. Both cycles involve energy being stored and then released into the environment. The act of the demon opening the door is ignored as an energy cost; likewise, the neuron's computation does not include the cost of communication. Each q$_i$ is a sample and represents the charge accumulated on the plasma membrane when synapse $i$ is activated.
}
    %Maxwell's demon cycle is analogous to the neuron's computational cycle.  The initial state in the demon cycle is equivalent to the neuron at rest. The demon sensing a fast molecules is analogous to the synaptic activations received by the neuron. Whereas the demon uses energy to set the memory and then opens the door for the molecule, the neuron stores charge on the membrane capacitance (C$_{\textrm{m}}$) and then pulses out once this voltage reaches threshold. Simultaneous with such outputs, both cycles then reset to their initial states and begin again. Both cycles involve energy being stored and then released into the environment. The act of the demon opening the door is ignored as an energy cost; likewise, the neuron's computation does not include the cost of communication. Each q$_i$ is a sample and represents the charge accumulated on the plasma membrane when synapse $i$ is activated.
    \label{demonvsneuroncycle}
  \label{demonfigure}
\end{figure}

To give this derivation a neural-like flavor, suppose a perfect integrator with the total synaptic input building up on the neuron's capacitance. Every so often the neuron signals this voltage and resets to its resting potential. Call the signal $ V_{sig}$, and rather unlike a neuron, let it have mean value (resting potential) of zero. That is, let it be normally distributed $\mathcal{N}(0, \sigma^2_{sig}=E[V_{sig}^2])$. The thermal noise voltage-fluctuation is also a zero-centered normal distribution, $\mathcal{N}(0,\sigma^2_{noise})$. Expressing this noise as energy on the membrane capacitance, {\large $\frac{C_m \sigma^2_{noise}}{2}=\frac{k\mathcal{T}}{2}\Rightarrow 
\sigma^2_{noise}=\frac{k\mathcal{T}}{C_m}$}
\cite{middleton, papoulis,carvermead}. 
Then using Shannon's result, e.g., theorem  10.1.1 as in \cite{cover}, the nats per transmission are
{\large $\frac{1}{2}\ln(1+\frac{\sigma^2_{sig}}{\sigma^2_{noise}})=\frac{1}{2}\ln(1+\frac{ C_m E[V_{sig}^2]}{k\mathcal{T}})$}
(with natural logarithms being used since we are performing a maximization, thus nats $=$ bits$\cdot \ln2$). Converting to bits, and calling this result the mutual information channel capacity,
{\large $C_{MI}=(2 \ln2)^{-1} \ln(1+\frac { C_m E[V_{sig}^2]}{k\mathcal{T}})$}. 

Next we need the energy cost, the average signal J/transmission developed on the fixed $C_m$ by the synaptic activation,
{\large $\mathcal{E}:=\frac {C_m E[V_{sig}^2]} { 2}$}. Dividing the bits/sec $C_{MI}$ by the J/sec $\mathcal{E}$ yields the bits/J form of interest; 
 {\large $ \frac{C_{MI}}{\mathcal{E}}=(C_m E[V_{sig}^2]\ln2 )^{-1}  \ln(1+\frac {C_m E[V_{sig}^2]}{k\mathcal{T}})$.  This ratio is recognized as the  monotonically decreasing function $\frac{\ln(1+x)}{cx} \;$with $x,c>0$}.  
Therefore maximizing over $E[V_{sig}^2]$ but with the restriction $E[V_{sig}^2]>0$, this is a limit result implying an approach to zero bits/sec. That is, \\
\noindent {\large $\underset{E[V_{sig}^2]\to 0}\lim \frac{C_{MI}}{\mathcal{E}}
=(C_m E[V_{sig}^2]\ln2 )^{-1}\frac {C_m E[V_{sig}^2]}{k\mathcal{T}}\\
\hspace*{1.9cm} =(k\mathcal{T}\ln2 )^{-1} \approx 1.6\cdot 10^{20} $} bits/J.

Two comments seem germane. First, physicists arrived at this value decades ago in their vanquishing of Maxwell's demon and its unsettling ability to create usable energy from randomness \cite{landauer61}. In their problem, the device (the demon) is not obviously computational in the neural sense; the demon just repeatedly (i) senses, (ii) stores, and (iii) operates a door based on the stored information, and then (iv) erases its stored information as it continues to separate fast molecules from the slower ones \cite{leff2002maxwell,sagawa}: see Fig \ref{demonvsneuroncycle}. Moreover, even after simplifying this cycle to steps (i), (ii) and (iv), physicists do see the demon's relevance to digital computation. Such a cycle is at the heart of modern computers where computation occurs through repetitive uses, or pairwise uses, of the  read/write/erase cycles. For example, bit-shifting as it underlies multiplication and the pairwise sensing and bit-setting (then resetting) of binary, Boolean logical operations reflect such cycles. Thus, as is well known from other arguments e.g., \cite{landauer61}, \cite{bennett82}, the limit-result of physics sets the energy-constraining bound on non-reversible digital computation. Regarding (iii) it would seem that if the demon communicates and controls the door as slowly as possible (i.e, the limit of time going to infinity), there is no need to assign an energy-cost to these functions.

In spite of a non-surprising qualitative comparison, there is a second insight. Compared to the estimates here of a neuron cycling from reset to firing to reset, this physics result is unimaginably more efficient, not just five or ten times more, but $10^8$-fold more efficient. Suppose that the computational portion of a human cortical neuron has capacitance  $C_m \approx 750$ pF (obtained by assuming the human neuron's surface area is  ca. three times a rat's pyramidal value of 260 pF \cite{singh2017consensus}) and suppose this neuron resets to $V_{rst}= -0.066 $ V while firing threshold is $V_\theta = -0.050 $ V. Then in the absence of inhibition, the excitatory synaptic energy needed to bring a neuron from reset to threshold is  
{\large $\frac {1}{2} C_m (V_{rst}^2-V_{\theta}^2) \approx 1.4 \cdot 10^{-12} $} J/spike. Assuming 4 bits/spike, the bits/J are $2.9 \cdot 10^{12}$. Compared to the optimal limit set by physics, this efficiency value is $10^{8}$ times less energy-efficient, a seemingly horrendous energy-efficiency for a supposedly optimized system.
 
\subsubsection*{The disagreement reorients our thinking}In the context of understanding neural computation via optimized energy-use, this huge discrepancy might discourage any further comparison with thermal physics or the use of mutual information. It could even discourage the assumption that Nature microscopically optimizes bits/J. But let us not give up so quickly. Note that the analogy between the four-step demon versus an abstract description of neural computation for one interpulse interval (IPI) is reasonable (see Fig \ref{demonvsneuroncycle}).  That is, (i) excitatory synaptic events are the analog of sensing,  these successive events are (ii) stored as charge on the plasma membrane capacitance until threshold is reached, at which point (iii) a pulse-out occurs, and then (iv)  the "memory" on this capacitor is reset and the cycle begins anew. Nevertheless, the analogy has its weak spots.

%END insert of lines 134-161
% INSERT 161 - 243.......................
The disharmony between the physical and biological perspectives arises from the physical simplifications that time is irrelevant  and that step (iii) is cost-free. While the physical simplifications ignore costs associated with step (iii), biology must pay for communication at this stage. That is, physics only looks at each computational element as a solitary individual, performing but a single operation. There is no consideration that each neuron participates in a large network or even that a logical gate must communicate its inference in a digital computer in a timely manner. Unlike idealized physics, Nature cannot afford to ignore the energy requirements arising from communication and time constraints that are fundamental network considerations \cite{laughlin2003communication} and fundamental to survival itself (especially time \cite{sterlingvijay,perge2012,perge2012}).

According to the energy audit, the costs of communication between neurons outweighs computational costs. Moreover, this relatively large communication expense further motivates the assumption of energy-efficient IPI-codes (i.e., making a large cost as small as possible is a sensible evolutionary prioritization). Thus the output variable of computation is assumed to be the IPI, or equivalently, the spike generation that is the time-mark of the IPIs endpoint.

Furthermore, any large energy cost of communication sensibly constrains energy allocated to computation.  Recalling our optimal limit with asymptotically zero bits/sec, it is unsustainable for a neuron to communicate minuscule fractions of a bit with each pulse out. To communicate the maximal bits/spike at low bits/sec leads to extreme communication costs because every halving of bits/sec requires at least a doubling of the number of neurons to maintain total bits/sec. Such increasing neuron numbers moves neurons farther away from each other (see SI Appendix), requiring longer axons to reach these other neurons and wider axons to avoid increased time delays, since such delays undermine the timely delivery of information \cite{sterlingvijay,perge2012}. This space problem arising from a larger number of neurons is recognized as severely constraining brain evolution and development as well as impacting energy-use \cite{mitchison92,stevensnips00,sejnowski_zhang2000universal,bullmore_sporns_2012economy,Karbowski15,clandininwang2016influence}. It is better for overall energy consumption and efficiency to compute at a larger, computationally inefficient bits/IPI that will feed the axons at some requisite bits/sec, keeping neuron number at some optimal level. To say it another way, a myopic bits/J optimization can lead to a nonsense result, such as zero bits/sec and asymptotically an infinite number of neurons.

\textit{Nevertheless, assuming efficient communication rates and timely delivery that go hand-in-hand with the observed communication costs, there is still reason to expect that neuronal computation is as energy-efficient as possible in supplying the required bits/sec of information to the axons}. The problem then is to identify such a computation together with its bits/J dependence and its inferred bits/sec.

\subsection*{A neurally relevant optimization} 
%\quad\\ 
%\FloatBarrier
\subsubsection*{How close is the optimized $N$ bits/J to 2500?}
\begin{figure}[t!]
\centering
\adjincludegraphics[width=8.8cm,Clip={.05\width} {0.007\height} {0.0\width} {.00\height}]{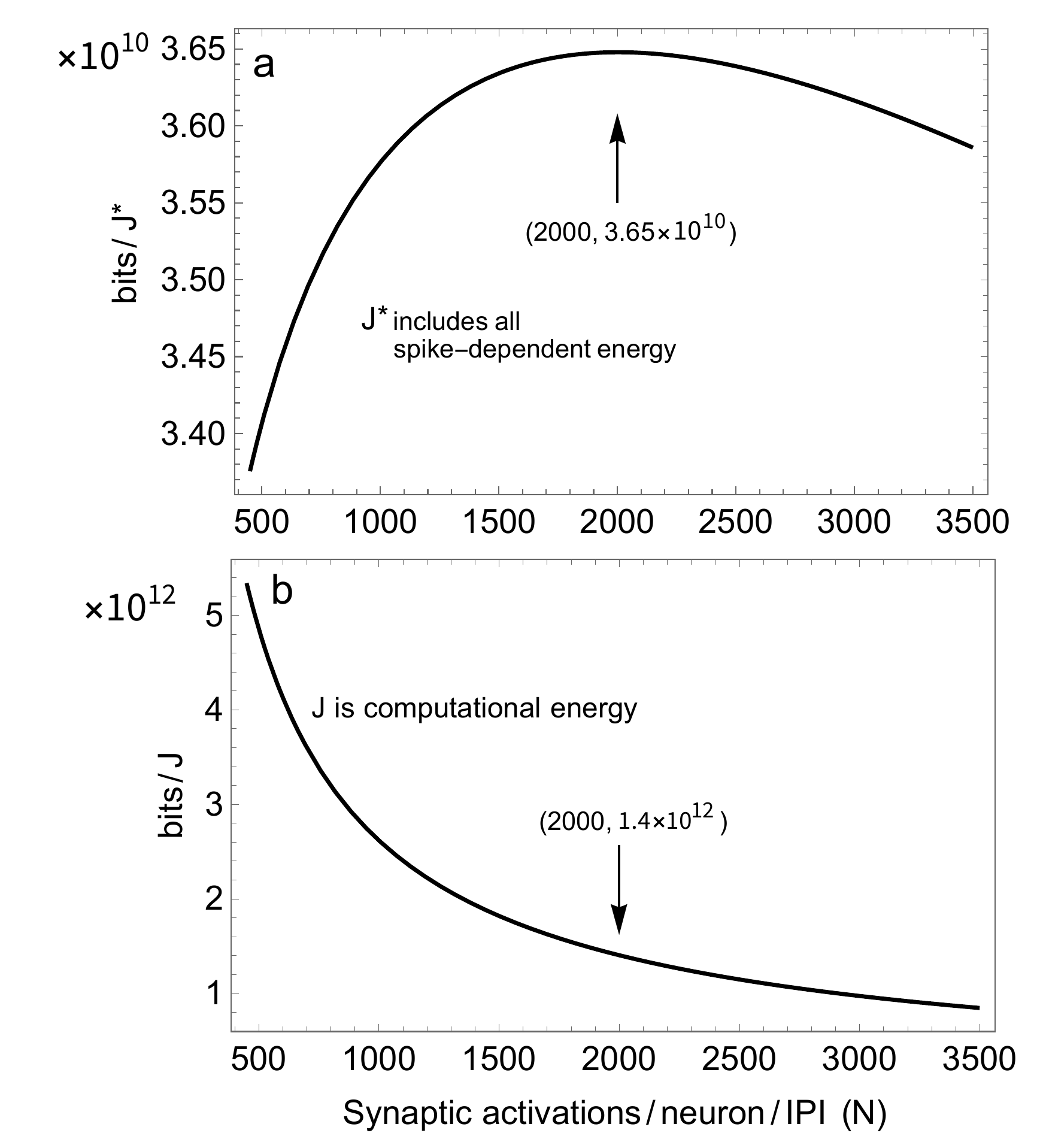}
\caption{Bits/J/neuron at optimal $N$.
(a) The bits/J function, Eq (1) accounting for  all spike-dependent energy-use,  is concave and reaches a maximum when $N$ is ca. 2000. This efficiency decreases little more than 5\% over a seven-fold range away from this 2000. At this optimum there are 7.48 bits/spike. (b) The optimal $N$ implies  $1.4 \cdot 10^{12}$ bits/computational-joule. (b) is calculated by changing Eq (1)'s denominator to $N\cdot B *E[T] \div (2500 \cdot n)$ instead of $J^*:=(A+N\cdot B/2500) *E[T]\div n$ of Eq (1).}
\label{optimalN}
\end{figure}

The computations of this section combined with the earlier energy-audit imply an efficiency of  ca. $1.4 \cdot 10^{12}$ bits/computational-J and less than 7.5 bits/IPI for neurons completing their first IPI. Comparing the curves of Fig \ref{optimalN}, the bits/J maximization that accounts for all spike-dependent costs produces the agreeable result $N \approx 2000$, which is not far from the 2500 derived earlier. By comparison, the purely computational perspective of costs, Fig \ref{optimalN}b,  indicates the exponentially increasing efficiency is  reached as a limit  $N \rightarrow 0$. \ref{optimalN}a also indicates  the optimization result is robust around the optimizing $N$, changing little over a 7-fold range;  likewise, bits/IPI is robust.  In sum for $N=2000$, the neuron computational efficiency is inferior to the demon by ca. $10^8$ but is optimal when other costs are considered. In fact, more detailed considerations below suggest slightly downgrading bit-rate estimates.

Using the notation $\Lambda$ for the random variable (RV) of the total, unfailed input intensity (events/sec) to a neuron and $\hat{\Lambda}$ the RV that is the neuron's estimate,
Fig \ref{optimalN}A illustrates the concave function being being maximized, {\large $\frac{I(\Lambda;T)}{\mathcal{E}(\Lambda,T)}
=\frac{I(\Lambda;\hat{\Lambda})}{\mathcal{E}(\Lambda,T)}
=$}  
$\dfrac %NUMERATOR
{\log_2(\ln(\dfrac{\hat{\lambda}_{mx}}{\hat{\lambda}_{mn}}))+\frac{1}{2}\log_2(\dfrac{(N+1)^2}{N})-\frac{1}{2}\log_2(2\pi e)} %DENOMINATOR FOLLOWS
{(A + N\cdot B/2500)
\cdot E[T]\div n}   \quad  (1) $ \\
where 
$I(\Lambda;T)$ and its equivalent $I(\Lambda;\hat{\Lambda})$ is the bits/IPI of information gain \cite{levybergersungkar}. This gain arises from  an additive neuron communicating its implicitly estimated latent variable's  value  $\hat{\Lambda}=\hat{\lambda}$ as a first-hitting time, $T=t$ (i.e., RV $T$ producing one particular realization, $t$). The denominator, previously introduced at the end of the energy audit, is the J/IPI per neuron as a function $N$. The ratio
$\dfrac{\hat{\lambda}_{mx}}{\hat{\lambda}_{mn}}$ is essentially the ratio of the maximum rate of synaptic activation  to the baseline rate.  

$I(\Lambda;T)$, derived in \cite{levybergersungkar}, requires Corollary 2 (below) for conversion to I($\Lambda;\hat{\Lambda}$). Moreover, attending this result are the related results, Lemma 2b and Corollary 1, which enhance our  understanding of the neuron's computation. Just before these mathematical developments, we recall and interpret some results of \cite{levybergersungkar}, one of which sheds light on the $10^8$ discrepancy with the demon-result. 

\subsubsection*{Deeper insights into the defined computation}
As noted in the introduction and developed in detail elsewhere \cite{levybergersungkar}, the neuron's computation is an estimation of its scalar latent variable $\Lambda=\lambda$.
$I(\Lambda ;T)=E_{\Lambda,T}[\frac {p(T|\Lambda)}{p(T)}]$ is the information gain for a Bayesian performing estimation \cite{lindley}. Written this way, the  relative entropy starts with the prior $p(t)$ and via sampling, i.e., synaptic activations, implicitly arrives at a likelihood $p(t|\lambda)$. The form of this conditional probability density is a maximum entropy development, which is the best distribution in the sense of maximizing a gaming strategy \cite{grunwald01}. The maximum entropy constraints are energy and unbiasedness. This likelihood also carries all of the information of sampling.

Defining (i) $\theta$ as threshold to fire, (ii) $E[V_{syn}]$ as the average size of the synaptic event arriving at the initial segment, and (iii) $E[V_{syn}^2]$ as its second moment, from equations 12 and 6 of \cite{levybergersungkar}, 
 $p(t|\lambda)=$ \\
{\large$\frac{\theta}{\sqrt{\pi \lambda t^3E[V^2_{syn}|\lambda]}} 
\exp(2 \frac {\theta E[V_{syn}| \lambda]} {E[V_{syn}^2|\lambda]}-\frac{\lambda t E[V_{syn}|\lambda]^2}{E[V_{syn}^2|\lambda]}-\frac{\theta^2}{\lambda t E[V_{syn}^2|\lambda]})$}.
\vspace{0.05cm}\\
While  the only consistent marginal distribution we have yet to discover is $p(\lambda)=(\lambda \log(\frac{\lambda_{mx}}{\lambda_{mn}}))^{-1}$ with $0<\lambda_{mn}<\lambda<\lambda_{mx}<\infty$, which is enough to infer the form of $p(t)$ and of $p(\lambda|t)$.
 
Importantly, the IPI, $t$, is a sufficient statistic, which is information-equivalent to the likelihood $p(t|\lambda)$ and so is the latent RV estimate,  {\large $\hat{\lambda}=\frac{N^2}{(N+1)t}$}. The conditional mean squared error of the estimate  is  {\large $E[(\hat{\Lambda}-\Lambda)^2|\lambda]
=\frac {\lambda^2(N+2)}{(N+1)^2}$} as Corollary 1 here demonstrates. Thus we not only define a neuron's computation, but can understand its performance as a statistical inference.

Parsing Eq (1), the information rate increases at the rate of ca. $\frac{1}{2}\log_2(N)$ while energy consumption increases in proportion to $N$. This disadvantageous ratio and the large optimizing $N$ help explain the demon's superior efficiency. Moreover, increasing the non-computational demands such that $A\div B$ increases, leads to a larger optimal value of $N$ and vice versa. Regardless, the corollaries of the next subsection clearly show that the energy devoted to computation, or other $N$-dependent energy consumers, restricts the precision of a neuron's estimation and restricts the information a neuron generates  when the neuron is required to be energy optimal.

\subsubsection*{Mathematical derivations}  As an approximation of a result in \cite{medallaLeubke}, assume an empirical distribution of synaptic weights such that the second non-central moment is equal to twice the mean squared (e.g., an exponential distribution). Note also that $\theta$ can be written as the product $N$, the average number of synaptic increments, multiplied by the average synaptic incrementing event $E[V_{syn}|\lambda]$ (with inhibition and capacitance taken into account, \cite{levybergersungkar}). That is, $\theta=N\cdot E[V_{syn}|\lambda]$. Putting this assumption to work, we obtain a simplification, and there are two new corollaries based on the above $p(t|\lambda)$.\vspace*{1mm}\\
\noindent$\textit{Lemma 1}$. {\large $p(t |\lambda) = N(2\pi \lambda t^3)^{-1/2} \exp(-\frac {\lambda t}{2}-\frac{N^2}{2 \lambda t} + N)$}.\vspace*{1mm}\\
\noindent $\textit{Proof}$: Start with $p(t |\lambda)$  given earlier, substitute using  
$\theta=N\cdot E[V_{syn}|\lambda]$, and then note that $\frac {E[V_{syn}|\lambda]^2}{ E[V_{syn}^2|\lambda]} =\frac {1}{2}$.
 
At this point there is an instructive and eventually simplifying transform to create $p(\hat{\lambda} |\lambda)$ from  $p(t |\lambda)$  The transform arises from the unbiased requirement, one of the constraints producing the earlier optimization results \cite{levybergersungkar}. As a guess suppose the unbiased estimate is {\large$\hat{\lambda}=\frac{N^2}{(N+1)t}$} or equivalently {\large$t=\frac{N^2}{(N+1)\hat{\lambda}}$}, then use this relation to transform $p(t|\lambda)$ to $p(\hat{\lambda}|\lambda)$.\vspace*{1mm}\\
\noindent$\textit{Lemma 2a}$: {\large $p(\hat{\lambda}|\lambda)=\\
\sqrt{N+1} (2 \pi \lambda \hat{\lambda})^{-1/2}
\exp(-\frac{\lambda N^2}{2(N+1)\hat{\lambda}} - \frac{\hat{\lambda}(N+1)}{2\lambda}+N)$.\vspace*{1mm}\\
\noindent$\textit{Lemma 2b}$: $E[\hat{\Lambda}|\lambda]=\lambda=\frac{N^2}{N+1}\cdot E[T^{-1}|\lambda]$};\\

\noindent So {\large$\hat{\lambda}=\frac{N^2}{(N+1)t}$} is indeed the desired unbiased  estimate, which has a particular mean squared error.\vspace*{1mm}\\
\textit{Corollary 1}: {\large$E[(\hat{\Lambda}-\Lambda)^2|\lambda]
=\frac {\lambda^2(N+2)}{(N+1)^2}$}.\vspace*{1mm}\\
$\textit{Proofs}$. See Methods. 

As the corollary shows, devoting more energy to computation by increasing $N$ reduces the error of the estimate. Specifically, the standard deviation decreases at the rate of $1/\sqrt{N}$. Of course, computational costs increase in direct proportion to $N$.

This corollary adds additional perspective to our definition of a neuron's computation as an estimate. Furthermore, the new likelihood, $p(\hat{\lambda}|\lambda)$, is particularly convenient for calculating information rates, a calculation which requires one more result. That result is the marginal distribution of $\hat{\Lambda}$. Because the only known sufficient density (and arguably the simplest) is
{\large $p(\lambda)=(\lambda \log(\frac{\lambda_{mx}}{\lambda_{mn}}))^{-1}$}, 
the estimate's marginal density is simply approximated via \vspace*{1mm}\\
%\noindent
\textit{Lemma 3.}
{\large$p(\hat{\lambda})=\int_{\lambda{mn}}^{\lambda{mx}} p(\lambda)p(\hat{\lambda}|\lambda)d\lambda    \approx 
(\hat{\lambda} \ln(\frac{\lambda_{mx}}{\lambda_{mn}}))^{-1}$}, \vspace*{1.5mm}\\
where the approximation arises by the near identity of the integral to $p(\lambda)$ assuming the range of $\lambda$ and $\hat{\lambda}$ is the same. Moreover, the lack of $\hat{\lambda}$ bias for all conditioning values of $\lambda$ hints that the approximation should be good. In fact using Mathematica at its default precision, numerical evaluation  of $\int_{\lambda{mn}}^{\lambda{mx}} p(\lambda)p(\hat{\lambda}|\lambda)d\lambda$ indicates zero difference between this integral and $(\hat{\lambda} \ln(\frac{\lambda_{mx}}{\lambda_{mn}}))^{-1}$.

The information rate per first-IPI can now be evaluated.\\
%\noindent
$\textit{Corollary 2}$. {\large $E_{T,\Lambda}[\log_2 \frac {p(T|\Lambda)} {p(T)}]=E_{\hat{\Lambda},\Lambda}[\log_2\frac {p(\hat{\Lambda}|\Lambda)} {p(\hat{\Lambda})}] \\
= \log_2(\ln(\frac{\hat{\lambda}_{mx}}{\hat{\lambda}_{mn}}))+\frac{1}{2}\log_2(\frac{(N+1)^2}{2\pi eN}) +\frac{1}{2}E_{\hat{\Lambda},\Lambda}[\log_2(\frac{\hat{\Lambda}}{\Lambda})]\\
\hspace{1.97 cm} \approx
\log_2(\ln(\frac{\lambda_{mx}}{\lambda_{mn}}))+\frac{1}{2}\log_2(\frac{(N+1)^2}{2\pi e N})$}.\\

\noindent$\textit{Proof}$: $E_{\hat{\Lambda},\Lambda}[\log_2 \frac {p(\hat{\Lambda}|\Lambda)}{p(\hat{\Lambda})}]=h(\hat{\Lambda})-h(\hat{\Lambda}|\Lambda)\approx h(\Lambda)-h(\hat{\Lambda}|\Lambda)$.

\subsubsection*{Limitations on the information rate} 
The bit/rate calculated above is arguably naive, even beyond the fact that we are assuming there is such a thing as an average neuron. First, under physiological conditions, humans are constantly making decisions, including novel sensory acquisitions  (e.g., a saccade and new fixation). Suppose  that such a decision-making interval and sensory reacquisition occurs every second. Then, many neurons do not complete even their first IPI. Such neurons make a much smaller information contribution, although still positive. To maintain average firing rate, suppose half the time a neuron completes one IPI, one-quarter of the time two IPIs, one-eighth of the time three IPIs, etc per decision-making interval. Thus half the time a neuron does not complete a first IPI, one-quarter of the time a neuron completes a first IPI but not a second, etc. Each non-completed IPI has a bit value. Combining the contributions for complete and incomplete IPIs produces a bit value of 5.1 bits/second for a 1 Hz neuron. See Methods for details.

Shot-noise is potentially deleterious to bit rate as well. As a crude approximation of  shot-noise affecting the signal, suppose Shannon’s independent additive Gaussian channel; i.e, the mutual information  is
$ \frac {1}{2} \log_2 \frac{\sigma^2_{signal} +\sigma^2_{noise}} {\sigma^2_{noise}}$. In biophysical simulations, depending on synaptic input intensity, it takes 50  to 250 Nav 1.6 activations to initiate an AP \cite{singh2017consensus}. Using this range as a Poisson noise and 2500 as the Poisson signal, the capacity is much smaller than the rate of information gain, 2.8  to 1.7  bits/sec. In fact simulations with this biophysical model produce  3 bits/IPI \cite{singh2017consensus}. This value is probably an underestimate by about one bit because the model did not contain inhibition; without inhibition, synaptic excitation rates are limited to less than 750 events to reach threshold vs the 2500 here allowed by inhibition and dendrosomatic surface area.

\section*{Discussion}
The Results contribute to our understanding of computation in the brain from the perspective of Nature. Essentially, the Results analyze a defined form of neural computation that is (i) based on postsynaptic activation and that is (ii) a probabilistic inference \cite{levybergersungkar}. From this defined perspective, the corresponding bits/J is maximized as a function of $N$. This value of $N$ is 2000, close enough to 2500 to substantiate the latter's  use in the audit. Likewise, it only changes the estimated synapses per neuron from 10000 to 8000 given a 75\% failure rate.

As first introduced into neuroscience in \cite{levybaxter96} and later emphasized by \cite{balasubramanian2015heterogeneity}, a certain class of bits/J optimizations can proceed if the denominator joule-term consists of two parts: a constant joule-consumption term added to a term in which the joule-consumption depends on the variable being optimized. Typically, this denominator consists of a constant energy-consumption term plus a firing-rate dependent term. However, here both denominator terms are dependent on the mean firing-rate (a constant); in addition, the second term is also dependent on $N$. Thus there is the $N$-based optimization of Fig \ref{optimalN}a while in Fig \ref{optimalN}b, the computational-bits/computation-J has no biologically meaningful maximization. Rather, \ref{optimalN}b notes the $1.4 \cdot 10^{12}$ computational-bits/computational-J at the optimal $N=2000$  from \ref{optimalN}a, where all relevant costs are included.

Another quantitative accomplishment is quantifying the $10^8$ discrepancy between the Demon's optimal computation and a neuron's optimal computation. From Fig 4b, we see that if Nature selects for smaller $N$, the efficiency increases exponentially. Indeed, a two-synapse neuron with 10,000-fold less surface area and a 1,000-fold decrease in the voltage between reset and threshold only misses the Demon's value by $10$-fold. However, using such neurons leads to other, larger cost increases if communication time is to remain constant. For a particular semi-quantitative analysis establishing this point, see SI Appendix. 

One reason why the bits/J/spike of Eq. 1 (Fig \ref{optimalN}a) increases so slowly as $N$ increase  is that the synaptic inputs are assumed to be  unclocked, asynchronous, and therefore approximately Poissonian \cite{bergerlevy10}. Unlike energy costs that grow in proportion to $N$, the  slow information growth at the rate $2^{-1}\ln(N)$ seems unavoidable \cite{vijayberry01}. Indeed  for visual sensing  \cite{niven07} notes a similar difference in growth rates. 

Although the basis of our theory is IPI coding, this hypothesis has some relevance to optimization theories based on rate-coding, e.g., \cite{vijayberry01,balasubramanian2015heterogeneity}. Specifically, each input to a neuron is a non-Poisson point process with an implicit rate. However, the union of these inputs is, to a good approximation, Poisson \cite{bergerlevy10}. This union of input lines creates the neuron's latent RV $\Lambda$. Thus, each neuron is estimating the intensity of a local-population rate-code over the time of each IPI. This may explain the similarity of bit-rate  estimates between models since as calculated in the Results and Methods, the randomness underlying this approximately Poisson signal is itself the largest source of uncertainty (i.e,  entropy). Finally, the rate-code approach (e.g., \cite{vijayberry01,balasubramanian2015heterogeneity}) might claim a greater generality as it applies to many pulses whereas the current IPI theory applies with exactitude only to the first IPI; the theory requires extensive work for application to later IPIs in cortex where neurons are receiving feedback during the generation of later IPIs, which feedback can change the value of the neuron's initial input.

\subsection*{Human and rodent energy audits}
The per-neuron values here are relatively close to those obtained by Herculano-Houzel \cite{Herculano_Houzel_2011}. Her value for the gray matter energy-use of human cortex is $1.32\cdot10^{-8}$ $\mu$mol of glucose per neuron per minute, which converts to $2.26\cdot10^{-10}$ W/neuron in terms of ATP. Our value is 1.94 $\cdot 10^{-10}$  W/neuron (Table S3). This small, 16\% difference is not surprising since she uses the older glucose values of slightly more than 20 W per brain, and we use her regional brain weight values and cell counts.
 
The top-down part of the audit can do no more than limit the total ATP available among the defined uses of ATP. This value is then subject to partitioning across specific, functional consumers.

%Except for this contribution, the top-down calculations are of no use in calculating computational energy use. The signal (computational energy) is smaller than the noise (variance of total energy available). The variance across the human population of the total energy available will, for the average cortical neuron of the average human,  always exceed the average energy expended by such a neuron for computation. Thus one must rely on bottom-up calculations, and here we look to the landmark work of Attwell and Laughlin \cite{attwell2001energy}. 

Staying as close to \cite{attwell2001energy} as sensible, newer values are used (e.g.,  for conversion of glucose to ATP \cite{nath2016thermodynamic} and  for the overlapping Na-K conductances of the AP \cite{hallermann2012state}). Species differences also create unavoidable discrepancies, including average firing rate, the fraction of the time that the glutamate-primed NMDARs are voltage-activated, and, more importantly, the surface area of rodent axons vs human axons. Other discrepancies arise from differences in partitioning of energy consumers. After removing WM costs, our partitioning of GM creates three subdivisions: computation, communication, and $SynMod^+$. Although the partitioning of energy consumption is at variance with  \cite{levybaxter96} and \cite{attwell2001energy}, this is not a problem because partitioning is allowed to suit the question. On the other hand, estimating the cost of $SynMod^+$ is problematic (see ouabain comments below). Moreover, the optimization here requires a subpartitioning of $SynMod^+$. The subcategories are partitioned as proportional to: time only, mean firing-rate, and mean firing-rate multiplied by $N$. Of relevance here,  the cost of synaptic modification, including metabotropic receptor activation and postsynaptically activated kinases do not fall within the present definition of computation but are activity-dependent $SynMod^+$ costs.

An earlier human GM energy audit \cite{lennie2003cost} comes to a different conclusion than the one found here. Although our more contemporary empirical values and more detailed analysis point to many initial disagreements with this study, these initial disagreements offset each other so that \cite{lennie2003cost}  concludes the GM has 3.36 ATP-W available, within 10\% of our 3.09. On the other hand, there are two rather important disagreements: (i) the postsynaptic current associated with the average presynaptic spike arrival and (ii) the total non-computational and non-communication energy expenditures which go hand-in-hand with a discrepancy of resting potential costs. Regarding (i), the relative postsynaptic currents per spike differ by nearly 14-fold, and this difference arises from three sources. First, in \cite{lennie2003cost} synaptic success rates are 2-fold greater than the rate used in \cite{attwell2001energy} and here. Second, the number of  synapses is 2.2-fold greater (we use newer values from normal tissue). Third, average synaptic conductance per presynaptic release is 3-fold greater than the values here (again using newer values \cite{medallaLeubke}). See Methods and SI Appendix for details. 

The other disagreement, (ii) arises because \cite{lennie2003cost} concludes that  50\% of the ATP goes to processes independent of electrical potentials (i.e., independent of computation plus communication). 
The earlier work bases its values on ouabain studies. While there is no argument that ouabain poisons the Na-K ATPase pump preventing it from metabolizing ATP, there is clear evidence that ouabain activates other functions known to increases ATP consumption. Ouabain increases spontaneous transmitter release \cite{nmj_ouabain1975} and depolarizes neurons. One must assume until shown otherwise that these two effects stimulate many ATPases and ATP-consuming processes that would not normally occur at rest. These include internal CaATPases to handle Ca$^+2$ influx \cite{ouabainCainflux}, vesicle recycling and transmitter packaging, metabotropic receptors, and possibly even synaptic modification that demands actin polymerization, membrane construction, protein insertion, and transport of the synthesized membrane to the ends of dendrites and axons.  In sum, any increases of ATP consumption will lead to underestimates of ATP used for the electrical potential and an over estimate of the ATP devoted to what we call $SynMod^+$.

Our ultimate problem with  \cite{lennie2003cost} are the two calculations of $N$ that it implies. Taking \cite{lennie2003cost}'s values of failure rate and synapse number implies that $N=8750$ while applying \cite{lennie2003cost}'s costs to the optimization of Eq. 1 produces $N\approx 300$. In contrast our two values are closer to agreeing, 2500 vs 2000.

%to  the information and biophysical calculations here create a startling contradiction. First, with so many more active synapses of greater conductance, the total charge injection to a neuron requires inhibition more than 2.5-fold greater than ours. This heavy downgrading implies average synaptic responses at the initial segment at the level of   thermal noise and a few-fold smaller than  shot-noise. This in turn implies huge information losses. Secondly, 

\subsection*{General relevance of Results}

\paragraph{Outside of neuroscience.}Because there is some interest e.g., \cite{ieeespectrum2017,neuromorphic_schuman2017survey} outside of neuroscience to reproduce neurally mediated cognition on a limited energy budget, the energy-audit here brings increased specificity to a comparison between the evolved biological vs the human engineered perspective. In particular, engineers often tout brain function as consuming energy at what they consider a modest 20 W given the difficulty they have in reproducing human cognition. Here we provide a more precise set of comparisons. Our computation can be compared to the job performed by the central processing unit. Communication has it's two major forms defined here, axonal costs and presynaptic functions, which must be compared to communication into and out of memories plus the communication of clock pulses. Perhaps maintenance can be compared to memory refresh costs. However, comparing power conversion loss by a computer to the heat generation of intermediary metabolism is challengeable since heating is fundamental to mammalian performance. A better comparison might be between the cost of cooling a computer and the biological heating cost.

\paragraph{Inside neuroscience.}Although the primary goal of the energy audit is  an estimate of the cost of computation \textit{per se}, the audit also illuminates the relative energetic costs of various neural functions. Notably for humans, the audit reveals that axonal resting potential costs are  greater than  the firing-rate costs. This axonal rest expense  is directly proportional  to the leak conductance and axonal surface area. Thus, of all the parameters, these two might benefit the most from better empirical data.  Regarding these large, leak-associated costs, two additional points seem relevant. First, regarding fMRI studies that measure regional brain metabolism, the small increases of oxygen consumption over baseline consumption \cite{foxraichle86} are consistent with the high, continuous cost of axonal  leak.

Second, arguing from her data and data of other studies \cite{Herculano_Houzel_2011}, Herculano-Houzel presents the intriguing hypothesis that average glucose consumption per cortical neuron per minute is constant across mammalian species. Qualitatively, this idea is consistent with the increase in neuron numbers along with the decrease of firing rates found in humans vs rats. However, it seems that the hypothesis can only be quantitatively correct if axonal leak-conductance in humans is much lower than in animals with smaller brains and presumably shorter axons of smaller diameters. This topic deserves more detailed exploration.

 Hopefully the work here motivates further empirical work, especially using primates, to improve the energy-audit and the calculations that ensue. Such empirical work includes better surface area measurements and a better idea about the NMDAR off-rate time constant. Finally, going beyond the average neuron, perhaps someday there will be energy-audits for the different cell types of cortex.
 
\matmethods{  

\subsection*{Partitioning glucose by region and by metabolic fate}

This section explains the top-down calculations of Table \ref{mainwatts}. The glucose-uptake values combine the regional uptakes, reported in terms of per 100 g of tissue from Graham et al. \cite{graham2002fdg} as copied into our Table S1
along with the reported regional masses from Azevedo et al. \cite{azevedo2009equal}. We choose this uptake study because of its use of the [$^{11}$C]glucose tracer and its straightforward application to obtain regional net glucose uptakes. Multiplying regional masses by uptake values, and converting to appropriate units as in Table S1, yields the first "Watts" column of Table \ref{mainwatts}. These glucose-watts are calculated using 2.8 MJ/mol \cite{nelson2008lehninger}. The regional uptakes are combined to produce the brain total as illustrated in Fig S1. 

Following the flow diagram of Fig S1, next we remove the non-oxidized glucose from regional and total uptakes. We use an oxygen-glucose index (OGI) value of 5.3 (out of 6 possible oxygen molecules per one glucose molecule). We assume the OGI is constant across regions and that we can ignore other, non-CO$_2$ carbons that enter and leave the brain. Thus, these simple glucose-watts are split into oxidized and non-oxidized as produced in Table \ref{mainwatts} and illustrated in Fig S1. 

As the energy source, the oxidized glucose is then partitioned into two different metabolic fates: heating and ATP. Again we assume this process is constant across regions and that the brain does not differ too much from other regions which have been studied in greater depth. The biological conversion is calculated using Nath's torsional mechanism, which yields 37 ATP molecules per molecule of glucose and 36,000 J/mol of ATP at 37$^\circ$ C. 

\subsection*{Computation Costs}
Our "on average" neuron begins at its reset voltage and then is driven to a threshold of -50 mV and then once again resets to its nominal resting potential of -66 mV. Between reset and threshold, the neuron is presumed to be under constant synaptic bombardment with its membrane potential, $V_m$, constantly changing. To simplify calculations, we work with an approximated average  $V_m,\;V_{ave}$ of -55 mV; this approximation assumes $V_m$ spends more time near threshold than reset. (Arguably the membrane potential near a synapse which is distant from the soma is a couple of mVs more depolarized than the somatic membrane voltage, but this is ignored.) To determine the cost of AMPAR computation, we use the ion preference ratios calculated from the reversal potential and use the total conductance to obtain a Na$^+$ conductance of 114.5 pS per 200pS AMPAR synapse as seen in Table S4. (The ion-preference ratios used for the calculations in Table S4 are calculated from the reported reversal potential value of -7 mV \cite{yoshimura1990amino} and the individual driving forces at this potential, $-90-(-7)=-83\,mV$ for K$^+$ and $55-(-7)=62\,mV$ for Na$^+$.) Multiplying the conductance by the difference between the Na$^+$ Nernst potential and the average membrane potential ($V_{Na,Nern}-V_{ave}$) yields a current of 12.5 pA per synapse. Multiplying this current by the SA duration converts the current to coulombs per synaptic activation, and dividing this by Faraday's constant gives us the moles of Na$^+$ that have entered per synaptic activation. Since 1 ATP molecule is required to pump out 3 Na$^+$ molecules, dividing by 3 and multiplying by the average neuron firing rate and success rate yields $1.29\cdot10^{-20}$ mols-ATP/synapse/sec. Multiplying by the total number of synapses  ($1.5\cdot10^{14}$) implies the rate of energy consumption is 0.069 W for AMPAR computation. When NMDARs are taken into account, the total computational cost is 0.10 W (assuming that NMDARs average conductance is  half as much as AMPAR's).

Table S4 lists the excitatory ion-fluxes mediated by AMPARs and NMDARs. The cost of the AMPAR ion fluxes is straightforward. The cost of NMDARs ion fluxes depends on the off-rate time constant as well as the average firing rate. That is, if this off-rate time constant is as slow as 200 msec and the IPI between firings of the postsynaptic neuron is 500 msec or more (such as the 1 sec interval that comes from the 1.0 Hz frequency used in the following calculations), then most glutamate-primed NMDARs will not be voltage activated. Thus, in contrast to the rat where the AMPAR and NMDAR fluxes are assumed to be equal, here we assume the ion-fluxes mediated by NMDARs are half that of the AMPARs and multiply the AMPAR cost by 1.5 to obtain the final values in Table S4. 

The spike-generator contributes both to computation and to communication; fortunately, its energetic cost is so small that it can be ignored.

\subsection*{Communication Costs}
Table S5 provides an overview of the communication calculations, which are broken down into Resting Potential Costs, Action Potential Costs, and Presynaptic Costs. The following sections explain these calculations, working towards greater and greater detail.

In general, the results for communication costs are built on less-than-ideal measurements requiring large extrapolations. For example, there does not seem to be any usable primate data. The proper way to determine surface area is with line-intersection counts, not point counts, and such counts require identification of almost all structures. As the reader will note in the supplement, use of mouse axon diameters produces much larger surface areas, thus raising communication costs and decreasing the energy available for computation and $SynMod^+$. 
%Likewise, copying recent values used in the biophysical literature for axon resting resistance (a rather difficult parameter to measure, especially for the small axons of interest here) also greatly increases the cost of communication compared to the values that we used \cite{hallermann2012state,hausser1997estimating}.

\subsubsection*{Resting Potential Costs}

The cost of the resting potential itself is simply viewed as the result of unequal but opposing Na$^+$ and K$^+$ conductances. If other ions contribute, we just assume that their energetic costs eventually translate into Na$^+$ and K$^+$ gradients. The axonal resting conductance uses the recent result of 50 k$\si{\ohm}$ cm$^2$ \cite{raastad2019slow}. With our surface area of $21.8\cdot 10^6$ cm$^2$ (includes axonal boutons, see Table S6), this produces a total conductance of  436 S. The driving voltage for each ion is determined by subtracting the appropriate Nernst potential from the assumed resting  membrane potential of -66 mV. Using Nernst potentials of +55 mV and -90 mV for Na$^+$ and K$^+$ resp., just assume  currents are equal and opposite at equilibrium.  Thus, conductance ratios derive from the equilibrium: $-24$ mV $\cdot g_K = -121$ mV $\cdot g_{Na}$; implying $g_K = 5.04\;g_{Na}$; further implying $\frac{g_{Na}}{g_{Na}+g_{K}}=\frac{1}{6.04}$. The Na$^+$-conductance times the driving voltage yields the  Na$^+$-current,  $0.121$ V $\cdot \frac{1}{6.04} \cdot 436$ S =  8.73 A. Scaling by Faraday's constant implies the total $Na^+$ influx; then divide by 3 to obtain  mols of ATP required to pump out this influx, 3.02 $\cdot 10^{-5}$ molATP/s. Multiplying by 36,000 J/molATP yields 1.09 W, the resting potential cost.

Plasma membrane leak is a major energy expenditure, 22\% of ATP-W here  compared to 13\% in \cite{attwell2001energy}. Here however, we emphasize that this cost is 66\% of gray matter communication costs. The differences in percentages arise from different interpretations of a functioning neuron and of the meaning of certain measurements. Our distinction between the cost of reset differs from their cost of resting potentials: here resting cost is entirely axonal and essentially continuous across time. Their resting cost is dendrosomatically based and deviates from our assumption that a neuron is under constant synaptic bombardment. 

\subsubsection*{Action Potential Costs}
Action potential costs are calculated from Na$^+$ pumping costs; see Table S5. The coulombs to charge a 110 mV action potential for the non-bouton axon starts with the product of the total GM axonal capacitance, $14.6$ F, the peak voltage, and the firing rate, 1 Hz; i.e., $14.6\cdot 0.11\cdot1.0 = 1.61$ amps. To account for the neutralized currents observed by Hallerman et al. \cite{hallermann2012state}, multiply this by 2.28, yielding 3.66 A. 

Bouton costs, although clearly part of an axon, are calculated separate from the axon. As will be detailed later, our approximation of surface areas treats all presynaptic structures as $bouton$ $terminaux$, and rather than assume tapering for impedance matching purposes, presume an abrupt transition of diameters. Importantly, we assume that a bouton mediates a calcium spike and that this spike only requires a 0.02 V depolarization to be activated. Altogether, the rate of $Na^+$ coulomb charging for boutons is $ 6.34$ F $\cdot 0.02$ V $\cdot1$ Hz = 0.13 A. 

The sum of axonal spike Na$^+$ and bouton charging determines the Na$^+$ to be pumped. Faraday's constant converts coulombs per sec to mols of charge per sec, yielding a Na$^+$ flux of $3.9\cdot 10^{-5}$ mol/sec. Dividing by 3 converts to ATP mol/sec; multiplying this value by Nath's 36,000 J/molATP yields the total action potential cost of 0.47 W.

 To calculate bits/J requires WMAP costs. Assume that the oligodendrocytes (especially myelogenesis) are using energy solely to support the AP. Then we approximate two-thirds of the 1.85 W energy goes to WMAP, $1.23$ W. 

The action potential values here largely agree with \cite{sterlingvijay}, but there are a number of important differences. They use an old, non-mammalian value for overlap. The neutralized current flux of the AP in mammals is  2.28 \cite{hallermann2012state} at the initial segment, far from the multiplier of four they use. Furthermore, the plotted values in \cite{sterlingvijay} Fig 7A are not adjusted for overlap. This figure uses an axonal length of 1 $\mu$m; therefore, for the axonal diameter plot-point of 0.5$\mu$m, the surface area is
$\pi/2\cdot 10^{-12}$ m$^2 = 1.57 \cdot 10^{-8}$ cm$^2$. This implies a capacitance of $1.57\cdot 10^{-14}$ F. Then the total charge needed for 0.1 V polarization is $1.57\cdot 10^{-15}$ coulombs. Multiplying by the number of charges per coulomb yields $1.57\cdot10^{-15}\cdot 6.24\cdot 10^{18}=9.8\cdot10^3\approx 10^4$ Na$^+$, the plotted value of fig 7A \cite{sterlingvijay}. Thus  the neutralized Na$^+$ flux was somehow lost  when the y-axis was labelled. With this understanding, our values only differ from \cite{sterlingvijay} because the calculations here use the mammalian measured overlap of 2.28. 

%(Remark: Howarth and Attwell \cite{howarth2012updated} use an overlap factor of 1.24 from the calculations of Carter and Bean \cite{carter2009sodium}, and this overlap factor is accurate for the soma according to Hallerman et al. \cite{hallermann2012state}. However, they show that the overlap factor is approximately 2.3-fold at the axonal initial segment. Thus, since the initial-segment action-potential shape is presumed to be more like the axon than the soma action potential, we use the 2.28 value.)  

\subsubsection*{Presynaptic AP Costs}
The presynaptic transmitter-associated costs are mostly based on the values of Attwell and Laughlin \cite{attwell2001energy} and of Howarth et al. \cite{howarth2012updated}. The assumptions include an assumed 25\% success rate of vesicular release for each cortical spike ($1.5\cdot 10^{14}$ spikes/sec under the 1 Hz and $1.5\cdot 10^{14}$ synapses assumptions). However, in contrast to Howarth et al. \cite{howarth2012updated}, which uses a number supported by observations in calyx of Held \cite{forsythe1998inactivation} and in cell cultures \cite{brody2000release}, the observations of Stevens and Wang \cite{stevens1995facilitation} in CA1 hippocampal pyramidal neurons indicate that the same calcium influx occurs for both synaptic successes and failures. Because adult hippocampal synapses seem a better model of cerebral cortical synapses than calyx or tissue culture synapses, we use the hippocampal observations. Therefore, the 1 Hz firing rate produces a Ca$^{2+}$ cost that is more than 8-fold greater than the cost of vesicle release events (VR events, Table S5). The Ca$^{2+}$ influx per action potential is  $1.2\cdot 10^4$ Ca$^{2+}$/vesicle, and assuming 1 ATP is required to pump out each Ca$^{2+}$, the Ca$^{2+}$ cost is $1.2\cdot 10^4$ ATPs/vesicle. Multiplying this by $1.5\cdot 10^{14}$ APs/sec for the gray matter, dividing by Avogadro's number, and finally multiplying by 36 kJ/molATP yields a total presynaptic Ca$^{2+}$ cost of 0.11 W. 

The cost per vesicle release is determined by adding the packaging and processing costs and then multiplying by the number of glutamate molecules per vesicle as in \cite{attwell2001energy} and  \cite{howarth2012updated}. Adding the cost of membrane fusion and endocytosis yields a total of 5,740 ATPs/vesicle \cite{howarth2012updated}. This value is multiplied by the VR events per second and divided by Avogadro's number to obtain $3.57\cdot 10^{-7}$ ATPmol/sec. Converting to watts yields a presynaptic transmitter release cost of 0.01 W and a total presynaptic cost of 0.12 W for the GM.

%\subsection*{Axonal and  presynaptic surface area}

%Surface areas of axons and their associated presynaptic structures are critical to the estimation of gray matter communication costs. Alas, the lack of  human data forces several bold extrapolations. Fortunately, some EM volume-fraction observations in other species and one well-quantified light microscopic (LM) study in cats help to constrain or serve as a check on  our assumptions.  

\subsubsection*{Synapse counts}
Both computation and communication costs depend on the number of cortical synapses. For the approach taken here, computational costs scale in a one-to-one ratio to synaptic counts while communication costs scale proportionally, but with a smaller proportionality constant.

The calculations use the Danish group's synapse counts of $1.5\cdot10^{14}$ \cite{pakkenberg2003aging}. The alternative to the numbers used here report an 80\% larger value \cite{alonso2008gender}; however, their human tissue comes from nominally non-epileptic tissue from severely epileptic patients. Since the incredibly epileptic tissue is likely to stimulate the nearby non-epileptic tissue at abnormally high firing rates, we find the data's import questionable. 
\subsubsection*{Estimation of Surface Areas from Mouse and Rabbit Data}

Here volume-fraction data are used to estimate axon and presynaptic surface areas. As far as we know, there are two journal-published, quantitative EM studies of cerebral cortex that are suitable for our purposes: one in rabbit \cite{schmolke1989structural} and one in mouse \cite{chklovskii2002wiring}. (Although structural identifications do not neatly conform to our simplifying cylindrical assumptions, we can still use their data to direct and to check our estimates.)
 
Chklovski et al. \cite{chklovskii2002wiring} report a 36\% volume-fraction for small axons, 15\% for boutons, 11\% for glia, 12\% for other, and 27\% for dendrites and spines as read from their graph in their Figure 3. They purposefully conducted their evaluations in tissue that lacked cell bodies and capillaries. Because cortical tissue does contain cell bodies and capillaries, this will produce a small error for the average cortical tissue. More worrisome is the size of "other," half of which could be very small axons. 
 
The quantification by Schmolke and Schleicher \cite{schmolke1989structural} examines the rabbit visual cortex. Their evaluation partitions cortex into two types of tissue: that with vertical dendritic bundling and that which lacks dendritic bundling (they do not seem to report the relative fraction of the two types of cortex, but we assume the tissue without bundling dominates over most of cortex). For boutons and axons respectively, they report volume fraction values within bundles of 17\% and 20\% and values between bundles of 26\% and 29\%.
 
The 30\% axonal volume fraction used in Table S6 is a compromise between the \cite{chklovskii2002wiring} value of 36\% and the two values from \cite{schmolke1989structural}. The average of the within bundle and between bundle volume-fractions from \cite{schmolke1989structural} is used for boutons.
Specifically, the approximated human volume fractions are (i) 22\% boutons, (ii) 30\% small axons, (iii) 11\% glia,  (iv) 5\% neuronal somata, (v) 3\% vasculature, (vi) 29\% dendrites, spineheads, and spine-stems, totaling 100\%. (It is assumed that standard fixation removes almost all of the physiological extracellular space and, naively, shrinkage/swelling has little relative effect on these values.) The calculations are essentially unaffected by the two conflicting bouton volume fractions since the difference between the two possible calculations is negligible. 

Table S6 lists the critical values, the intermediate values for the cylindrical model to fit the data, and finally the implications for the relevant membrane capacitance.  
\subsubsection*{Cylindrical model approximations for axons and boutons}

\textit{Axons}: By making a cylindrical assumption and assuming the average small axon's diameter is $0.50\;\mu $m (radius = $0.25\cdot10^{-4}$ cm), a small extrapolation of a cross-species result in the cerebellum \cite{wyatt2005speed}, we can estimate the total surface area of these unmyelinated axons using the 30\% volume-fraction to calculate the length of an average axon, $L_{ax}$. The total volume (cm\;$^3$) occupied by all such axons is 
$L_{ax}\cdot 1.5\cdot 10^{10}\cdot \pi (0.25\cdot 10^{-4})^2 $. Dividing this volume by the volume of the GM (632 cm$^3$) must equal the volume fraction, 0.3. Solving yields $L_{ax}=6.44$ cm. 
Then net surface area is calculated using this length, the same diameter and number of neurons, $6.44\cdot 1.5\cdot 10^{10}\cdot \pi \cdot 0.5\cdot 10^{-4}=1.52\cdot 10^7 $ cm$^2$. For an independent calculation of axon length based on LM data, see SI Appendix.

\textit{Boutons}: The surface area estimates also treat boutons (Btn) as uniform cylinders of a different diameter. Assume that cortical presynaptic structures in humans are no bigger than in any other mammalian species. To determine bouton surface area, assume a bouton diameter ($d_{pb}$)  $1.1 \;\mu m$ and height ($h_{pb}$) $1.0\; \mu m$. Denote the total number of synapses in the gray matter as $n_{gm}$  ($1.5 \cdot 10^{14}$). (Note that the cylinder area of interest has only one base.) Then, with the formulation $A_{pb} = n_{gm} \pi (d_{pb}h_{pb} + (\frac{1}{2} d_{pb})^2)$, the bouton surface area works out to $A_{pb} = 1.5 \cdot 10^{14} \pi (1.1\;\mu m \cdot 1.0\;\mu m + (0.55\;\mu m)^2) = 6.61\cdot 10^6$ cm$^2$. See Tables S6 and S7.

We assume a bouton only accounts for one synapse. However, larger boutons can contact multiple, distinct postsynaptic neurons. Thus the small cylinders, as individual synapses, are an attempt to approximate such presynaptic configurations. See Table S8 for more details and for the effect of overestimating areas.

\subsection*{Oxidized vs. non-oxidized glucose}
Arteriovenous blood differences indicate that insufficient oxygen is consumed to oxidize all the glucose that is taken up by the brain. Supposing glucose is the only energy-source, it takes six O$_2$'s for complete oxidation.  The calculations use an OGI value of 5.3 \cite{vaishnavi2010regional}. Other values from arteriovenous differences are found in the literature \cite{wahren1999brain,boyle1994diminished,rasmussen2010brain}. Even before these blood differences where observed, Raichle's lab  proposed as much as $20\%$ of the glucose is not oxidized \cite{fox1988nonoxidative}. 

\subsection*{Glucose to ATP based on Nath's theory}
Table S2 offers the reader a choice between  Nath's torsional conversion mechanism of glucose to ATP \cite{nath2016thermodynamic,nath2010beyond,nath2017two} versus the conventional conversion to ATP based on Mitchell's chemiosmotic theory \cite{mitchell1966chemiosmotic}.  According to Nath, the minimum number of ATP molecules produced per molecule of glucose oxidized is 32, and this includes mitochondrial leak and slip \cite{nath2016thermodynamic}. Nath's calculations are based on free-energy values under physiological conditions. However, his calculations are recent while the standard model has been taught for decades, although not without controversy \cite{villadsen2011thermodynamics}. The standard textbook number for this conversion is 33 ATPs per molecule of glucose before accounting for mitochondrial proton leak and slip. Since leak is often assumed to consume 20\% of the energy that might have gone to ATP production in oxidative phosphorylation \cite{attwell2001energy,rolfe1997cellular}, the Mitchell conversion number is reduced from 33 to 27 molecules of ATP (2 ATPs are produced by glycolysis and 2 by the Krebs cycle, so this 20\% reduction only applies to the ATP produced in the electron transport chain).

%The other choice given to the reader in Table S2 is the choice between two different firing rates. When the higher firing-rate or the Mitchell mechanism is used, there is no energy available for $SynMod^+$. Thus in these cases, the accounting cannot be balanced. In this regard, an energy-allocation for maintenance and synaptic modification ($SynMod^+$ in Table \ref{mainwatts} and 2) is a bare minimum and is just estimated via the guess that its value is equal to the computational cost. 

\subsection*{$SynMod^+$}
Here $SynMod^+$ is not directly calculated. Rather it is residual of the energy available after removing the above uses. The assumed subpartitioning occurs as follows. Assume 10\% of this goes to time proportional costs; assume the postsynaptic fraction, accounting for metabotropic activations, receptor modification, and actin polymerization-depolymerization cycles, equals $0.134$ W, which is activity and synapse number dependent. The remainder,  devoted to synaptogenesis and firing rate dependent axonal and dendritic growth (e.g., membrane construction, protein insertion, axo- and dendro-plasmic transport) is just activity dependent.

\subsection*{Proofs}
The proof of \textit{lemma 2a} is just a textbook change of variable from one density to another \cite{moodgraybill} where $dt=\frac{N^2}{(N+1)\hat{\lambda}^2}d\hat{\lambda}$; to prove \textit{corollary 1} and the first equality of \textit{lemma 2b}, use \textit{2a} to calculate the appropriate conditional moments, which Mathematica obliges; to prove the second equality of \textit{2b}, use lemma $1$ to calculate the indicated conditional moment.

\subsubsection*{Parameterizing the marginal prior $p(\lambda)$}
As derived from first principles \cite{levybergersungkar}, the only known, consistent marginal prior of the latent RV is $p(\lambda)=(\lambda \ln(\frac{\lambda_{mx}}{\lambda_{mn}}) )^{-1}$ where the bounds of the range of this RV, and thus its normalizing constant, are the subject of empirical observations and the required definition
$\lambda \in (0<\lambda_{mn} < \lambda_{mx}<\infty)$. 
%Recall that $\lambda$ is the rate of activations of $10^4$ input lines undergoing a 25$\%$ success rate when activated. Then on average $10^4$ is also the rate of activations if the neuron is to fire on average at 1 Hz. To achieve at minimum IPI of 2.5 msec requires, over a period of 2.5 msec, an activation rate of $10^6$ and by chance no synaptic failures. Substituting as $k:=\frac{mx}{mn}$ re-expresses the prior as  $p(\lambda)=(\lambda \ln(\frac{k}) )^{-1}$ with a mean of $E[\Lambda]=\frac{mn(k-1)}{Log[k]}$. Setting 

From the energy-audit, use the 1 Hz average firing rate. Then $E[\Lambda]$, the mean marginal total input firing rate, is $10^4$/sec. Now suppose that the rate of spontaneous release is 1 Hz over these $10^4$ synapses giving us  $\lambda_{mn}=1$.  With one unknown in one equation, 
$E[\Lambda]=\frac{\lambda_{mx}-\lambda_{mn}}{\ln(\frac{\lambda_{mx}}{1})}=10000$, Mathematica produces  $\lambda_{mx} \approx 116672$, and the prior is fully parameterized.  
%The sensitivity of the bit-rate calculations is revealed by keeping the mean firing rate constant while increasing or decreasing $\lambda_{mn}$ by ten-fold. When increased ten-fold, $\lambda_{mx}=32335.3$; when decreased ten-fold $\lambda_{mx}=52699.9$. The corresponding bit rate changes are -0.4  and +0.3 bits/sec/neuron. For details concerning concerning the ignorable approximation of the $\hat{\Lambda}$ distribution by that of $\Lambda$  see SI.

\subsubsection*{Adjusting the bit-rate calculation for multiple IPIs per decision-making interval (DMI)}
 The 7.48 bits per IPI only applies to a neuron's first IPI. Later spikes are worth considerably less using the current simplistic model of a fixed threshold and no feedback. Moreover, while maintaining the average firing rate, we might suppose that only half  the time a neuron completes a first IPI, half of these complete a second IPI, and so. Thus the average number of spikes per DMI remains nearly one. With  a fixed threshold, the bit values of the later spikes are quite small. The value of the second through fourth spikes are $\{\frac{1}{2}\log_2(\frac{2N}{N}),\;\frac{1}{2}\log_2(\frac{3N}{2N}),\;\frac{1}{2}\log_2(\frac{4N}{3N})\}$ which gives ca. 0.35 bits.  However complementing the completion of first IPI is, half the time, the bit contribution of an  uncompleted IPI, $0.5\cdot 1$ and for the one-quarter  of the time a neuron produces a first IPI but not a second, and so on for later IPIs. The summed value of these non-firings approaches 1 bit. Then,  $7.48/2 +0.35 +1\approx 5.1 $ bits.  
 \subsubsection*{Shot-noise can effect bit-rate but not as much as the signal}
As measured in the biophysical simulations \cite{singh2017consensus}, the most deleterious degradation of a neuron's computation arises, not from thermal noise or shot-noise \cite{laughlin2003communication}, but from the neuron's input signal itself. Here is a calculation consistent with this biophysical observation.
 
Using stochastic NaV 1.2 and NaV 1.6 channels in a biophysical model of a rat pyramidal neuron, it is possible to observe shot-noise and to estimate the number of such channels that are activated at threshold. With relatively slow depolarization, there are less than 250 channels on when threshold is reached, and this number of channels seems to contribute less than 1.6 mV (see Fig 5 in \cite{singh2017consensus}). Thus modeling channel activation as a Poisson process with rate 250 and individual amplitudes of 6.4 $\mu$V, Campbell's theorem \cite{parzen} produces the variance; this variance is less than  
$250 \cdot (6.4\cdot 10^{-6})^2=1.02 \cdot 10^{-8}$. The same calculation for the input excitation yields a variance of 
$2500 \cdot (6.4\cdot 10^{-6})^2=1.02 \cdot 10^{-7}$, a $10:1$ ratio.

%THIS IS REPLACED BY TEXT IN RESULTS:Then, the net variance is represented by multiplying the drift variance by something that increases it less than 10\%, say 1.09. This nine percent greater variance reduces the information gain by
%$\log(1.09)\approx 0.12 $ bits. 
\subsection*{Numerically-based optimization calculations}
Optimizing the bits/J equation uses Mathematica. Treat $N$, the average number of events per IPI, as a continuous variable.  Then to optimize, take the derivative, $dN$, of the single neuron, single IPI bit/J formulation. Set the numerator of this derivative equal to zero and solve for $N$ using Mathematica's NSolve.

}
\showmatmethods % Display the Materials and Methods section
\acknow{The authors are grateful for comments and suggestions of earlier versions provided by Costa Colbert, Robert Baxter, Sunil Nath, and David Attwell.}

\showacknow{} % Display the acknowledgments section

% Bibliography
\bibliography{pnas}

\section{SUPPLEMENTARY APPENDIX}
\setcounter{table}{0}
\renewcommand{\thetable}{S\arabic{table}}

\setcounter{figure}{0}
\renewcommand{\thefigure}{S\arabic{figure}}

\section*{The added expense arising from using neurons that approach Demon values in energy-efficiency of computation}
Consider the extreme construction which replaces the prototypical 750 pF neuron analyzed here with a system of  many  small, computationally energy-efficient neurons.
The replacement system is made of many miniature neurons; each miniature neuron has  just two synapses.  Such a miniature neuron  computes near the demon-inspired efficiency, kTln2,  because the surface area, and therefore the capacitance,  can be reduced about ten-thousand fold (or at least down to the limit not much bigger than the size of a mammalian cell nucleus); additionally, the voltage from reset to threshold can be reduced  1000-fold. This gives us a nominal $10^7$ savings in postsynaptic, computational energy, within $10^1$ of kTln2. However, we must use a system constructed of such miniature neurons to replace our conventional neuron because we assume that conventional neural computation must use learned information that requires combining $2^{13}=8192$ synapses (close to 10000 and powers of 2 are convenient here). To produce the needed combination of signals in this system of miniatures, one needs 8191 miniature neurons arranged in a hierarchy of 13 levels with $2^{12}$ miniature neurons at the bottom, with each succeeding level having half the number of neurons of its input level, until the combined signals reach one miniature neuron at the top. This system requires twice the original $2^{13}$ synapses, i.e., $2^{14}$ synapses ($2^{13}$ come into to the bottom level and then half again as much for each succeeding level). This extra $2^{13}$ synapses has the regular presynaptic costs which   wipes out the original postsynaptic savings that we got from the miniatures' rescaled capacitance rescaled voltage-range. (Presynaptic costs will be a little less than 0.12 W; see Table S5).  \\

More to the point, this system of 8191 neurons (call it a  surrogate neuron) takes up  more space. Assuming synapses  have a fixed size; then these pre- and postsynaptic structures occupy more than 1/3 of the space of whole brain. Additionally the volume of the 8191 miniature neurons must also include the cell bodies; assume a  nucleus size of 4 $\mu$m for each of these miniatures. Then together they occupy several fold more volume than all the dendrites of a regular neuron. And finally there is the volume occupied by the axons and presynaptic structures of the neuron at the top of the surrogate neuron. \\

This top-miniature neuron has an axon that replicates the connectivity of a standard axon. So in this model it has 8192 presynaptic structures. Clearly, the surrogate neuron is bigger than a regular neuron.\\

Now assume that communication time between regular neurons has evolved to fit the ecological niche of a human $10^4$ or more years ago. Then brain computational speed depends on matching this niche and cannot be allowed to vary. However due to the increased volume occupied by the surrogate neuron, the distance between surrogates is greater than the distance between regular neurons. Since each  output axon from the top-miniature neuron is equivalent to a conventional neuron’s axon, this surrogate axon must travel over a larger distance than the equivalent, regular-neuron axon. \\

Because  communication time between surrogate neurons cannot be allowed to increase from normal, the top-miniature-neuron's axon must not only be longer but it also must be wider.  Thus due to this axon's surface area increase, communication costs go up, and they have gone up far beyond the savings in computational costs.  

\section*{Selecting, adjusting, and commenting on literature values for glucose uptake }
In the literature, there are various uncertainties and incompatibilities which require compromises and approximations.
The subcortical distinctions made by Azevedo et al. \cite{azevedo2009equal} and Graham et al. \cite{graham2002fdg} are different. That is, the purely anatomical study lumps together the striatum, thalamus, colliculus, hypothalamus, pons and medulla while the reported [$^{11}$C]glucose measurements only include the individual subcortical regions caudate, putamen and thalamus. Assuming these subcortical forebrain regions account for much of the weight of what Table 1 and Table \ref{glupartitioning} labels as "other regions", and assuming that the unmeasured brain regions are not too different in glucose uptake, we just use a single value and multiply by the weights of the subcortical regions given by Azevedo et al. \cite{azevedo2009equal}. 

The [$^{11}$C]glucose uptake by the  choroid plexus and brain capillaries is assumed to be negligible. 
The ventricular weight is inferred from Azevedo et al. \cite{azevedo2009equal} to produce their total of 1510 g.

Another approximation is required due to the non-uniformity of brain size. The representative brain mass values available are for an average male brain. 
Incompatibly, the regional glucose uptake values from Graham et al. \cite{graham2002fdg} are averages from six females and four males with no statements about sex differences. Due to this sex heterogeneity there will be large total brain weight heterogeneity and a lower average brain mass per subject than the 1510 g value. Thus some scaling or conversion is needed. To remain consistent, the Graham et al. regional uptake values are scaled by the Azevedo et al. regional brain weights. Table \ref{glupartitioning} details these calculations. As a result, instead of a total brain uptake rate of 6.48 $\mu$ mol/sec the summed regional uptake rates yields 6.05 $\mu$ mol/sec. 

The OGI number is problematic in terms of accuracy and reproducibility. One group with multiple publications on the topic reports an OGI value of slightly more than 5.2 \cite{overgaard2012hypoxia}. Other A-V studies favor a higher OGI  (ca. 5.6, \cite{wahren1999brain}), but several of these studies also favor decreasing the value of glucose uptake by virtue of the net efflux of non-CO$_2$ carbons. Supposing the larger OGI is correct and supposing that we are allowed to ignore non-CO$_2$ carbon efflux, such an increase in oxidized carbons could yield an additional ca. 0.14 ATP-watt available to gray matter.

%CAN WE DELETE THIS SINCE WE DO PARTITION WM NOW FOR THE OPTIMIZATION???As Fig. 2 in the main text and Fig. \ref{partitioning} show, we do not bother to subdivide the energy-use of the white matter beyond its total ATP-watts. In contrast, gray matter is further subdivided. Specifically, separate bottom up calculations are used to determine communication and computational energy consumption (see below).

\section*{Conversion of glucose to ATP}
There are three problems with the textbook calculations and previous brain calculations of glucose to ATP conversion: such calculations are based on room temperature free-energies, the mitochondrial leak value is based on a maximal leak value from non-neuronal tissue, and the chemiosmotic hypothesis ignores slip in redox pumps. Slip would further reduce the chemiosmotic ATP values by 10\% (Nath, personal communication). However, Nath's novel mechanism does not require any ATP downgrades. (He identifies the neutral form of succinic acid as mediating leak. This form can penetrate the cristal membrane as the dianionic form, creating slip while the succinate monoanion is the motive form. Thus, both leak and slip are accounted for \cite{nath2016thermodynamic}.) Here, however, we do not attempt to account for the issue of slip in our chemiosmotic calculations. Thus, the choice offered for ATP production in the gray matter in Table \ref{mainwattsnathandmitchell} is 3.09 W (Nath) or 2.61 W (Mitchell). Both of these calculations use the conversion factor of $36$ kJ/molATP \cite{nath2016thermodynamic,nath2009energy} rather than the room-temperature value typically used. 

Although we consider Nath's torsional mechanism to be a more accurate depiction of ATP production than the chemiosmotic mechanism, we recognize that this newer mechanism is not directly informed by brain mitochondrial studies. That is, brain tissue has different uncoupling proteins which could potentially alter the amount of ATP produced per mole of glucose. Of course, different species have different issues concerning thermoregulation and thus may differ in amount of leak (see for example \cite{melanie2019allometry}). Such studies argue that larger animals are more efficient in regard to mitochondrial production of ATP. Thus the conversion values of glucose to ATP used for rats are plausibly lower than for humans. 

\section*{Other contrasting estimates with earlier results in the literature}
\subsubsection*{Postsynaptic ionotropic costs are different}
Our postsynaptic costs for AMPAR activation are about half of Attwell and Laughlin's ($79\cdot10^6$ ATP/action potential/neuron vs $134\cdot10^6$ ATP/action potential/neuron) \cite{attwell2001energy}. The difference mostly arises from their use of an outlier value for synaptic conductance that is doubted by the authors themselves as cited in Attwell and Laughlin. Biophysical simulations encourage us to discard the outlier value. 

The biophysical simulations in Singh and Levy \cite{singh2017consensus}, which use the consensus layer 5 prefrontal pyramidal neuron found in a variety of biophysical models, support the lower postsynaptic activation costs used here. Such simulations use 200 pS AMPAR-conductances per synapse and no NMDARs. Under these conditions, it typically takes 250 to 700 synaptic activations to fire the neuron. Such simulations did not include inhibition. Upgrading the model to include inhibition, as is done here and as occurs in the several articles that consider balanced inhibition (e.g., \cite{balancedinhibabbottannev,balancedinhibniven2013}), is required to be consistent with the estimate of 2500 synaptic activations per output spike. That is, inhibition is implicitly incorporated into our cost calculations by virtue of increasing the number of synaptic activations needed to reach threshold.

The same amount of excitatory synaptic activations (ca. 2500 input activations per pulse out) can be true for both humans and rats. That is, although the dendritic surface area of a human cortical neuron is greater than a rat's, a lower rate of inhibition in the human can compensate so that the same amount of excitatory synaptic activation is required to fire either neuron.

\subsection*{Sensitivity to axon and presynaptic assumptions} 
Gray matter communication costs are rather large and are directly a function of axonal and presynaptic surface areas. Therefore, it is worth revealing the sensitivity of the calculations to the assumptions going into these surface area calculations.

\textit{Axons}: 
Fixing the volume-fraction at 30\%, and varying axon diameter changes the values of surface area and axon length. Table \ref{SAandcomm} shows the relationship between these parameters as well as their affect on aspects of communication costs. Mouse data motivates much narrower axons but not smaller than 0.25 $\mu$m \cite{faisal2007stochastic}. As the smallest possible diameter, we choose the $0.28\;\mu m$ mouse-inspired diameter \cite{braitenberg1998cortex}. 

%Merely changing diameter from 0.5 to 0.4 $\mu$m increases the axonal contribution to total communication costs by 0.3 W.  Clearly the mouse diameter increases energy-use well beyond that which is available based on top-down calculations THIS ISN'T NECESSARILY TRUE NOW THAT WE USE A LOWER FIRING RATE AND THE 50K RESISTIVITY SO SHOULD WE JUST DELETE???. 

\textit{Boutons}: As noted before, the cylindrical assumption for boutons is crude. Table \ref{boutons} illustrates the sensitivity of our bouton size assumptions and, by extension, our bouton capacitance values relative to volume fractions. In these calculations, the different volume fractions are a necessary result of varying the bouton dimensions.

\subsubsection*{Reconciling the cost of \textit{$SynMod^+$}}
First, it must be said that the catchall partition labeled $SynMod^+$ has never been properly measured for an adult animal as far as we know. More to the point here however, is that our perspective on what $SynMod^+$ should include differs somewhat from earlier work.

What follows explains our rejection of previous estimations of the ATP-use attributed to the catchall and unpartitoned "$SynMod^+$" category of energy consumption. In the rodent audit, the energy consumed by $SynMod^+$ is based on published research that: (i) mathematically differences ATP-use before and after inhibition of the Na-K ATPase pump and that (ii) implicitly but necessarily assumes that removing the Na$^+$ gradient will not increase other forms of ATP-use. However, the literature argues otherwise. In general, cardiac glycosides such as ouabain, will accelerate several forms of ATP consumption. 

 Ouabain causes depolarization because of leak conductance, and it also causes transmitter release \cite{meyer1981correlations,santos1992effect,satoh1992mechanism,lomeo2003exocytotic} and in particular quantal transmitter release \cite{nmj_oubain1975}. Quantal transmitter release implies (i) the ATP-consuming processes of vesicle recycling and re-loading, (ii) postsynaptic activation of GTP-consuming metabotropic receptors, and (iii) postsynaptic activation of calcium-conducting NMDARs, which will activate various postsynaptic, ATP-consuming kinases. Moreover, in addition to the vesicle recycling, metabotropic, and kinase costs, there are both pre- and postsynaptic calcium pumping costs. That is, poisoning the Na-K pump leads to increase of [Ca$^{2+}$]$_{in}$ as does the NMDAR activation. Raising levels of internal calcium in turn activates one or more of at least three types of Ca$^{2+}$-ATPases: those in the plasma membrane, those associated with sarco- and endoplasmic reticula, and a mitochondrial accumulator \cite{mitocondrialreticulumParekh, calciumpumps2005, Ca_reticulumverkhratsky2005}. Any other kind of poisoning experiment that causes depolarization will produce similar increased demands for ATP. Finally, there are those (not Attwell et al.) that assume blocking action potentials will remove communication costs, but our leak calculations refute this idea.
 
 \subsection*{Comparison to a class of completely quantified axons}

Concerning axon lengths, there is one LM study that provides data indicating the reasonable nature of the length estimates obtained above. In particular, there are axonal length data for cat L2/3 pyramidal cells using 30 completely stained axons. These data imply an axonal length of 4 cm per neuron \cite{binzegger2010axonal}, but as LM measurements, they will incorporate terminal boutons. For comparison to the estimates here, we combine axonal boutons and the axonal lengths without terminal boutons.  Assume there are ten thousand boutons per neuron, each $1.1\cdot 10^{-4}$ cm in length. Using a volume fraction for small axons of exactly 30\%, our unadorned axon length is 6.44 cm per neuron. Adding the bouton lengths yields 7.44 cm per neuron. Thus we are predicting that LM quantification of the average human pyramidal neuron's axon is about 86\% longer than the cat axon. More details on the derivations and parametric sensitivities are found in Supplement including Tables S6-S8. 

\setcounter{figure}{0}
\setcounter{table}{0}

\section*{Probability and entropy approximations}
\subsection*{Initial development of approximations}
There are two approximations needed to calculate a valued Lindley-Shannon information rate,
$h(\hat{\Lambda})-h(\hat{\Lambda}|\Lambda)$. After an exact integration by Mathematica, the first approximation follows. Specifically,
$ \\ \noindent
p(\hat{\lambda})=
\int_{\lambda_{mn}}^{\lambda_{mx}} p(\lambda)p(\hat{\lambda}|\lambda)   d\lambda      = \\$
$( \hat{\lambda} \ln(\frac{\lambda_{mx}}{\lambda_{mn}}))^{-1} \cdot \frac{1}{2}
(\mathrm{erf}\left(\frac{N\lambda_{mx}    -(N+1) \hat{\lambda}}
{\sqrt{2(N+1)\hat{\lambda} \lambda_{mx}}}\right)
-\mathrm{erf} \left(\frac{N\lambda_{mn} +(N+1) \hat{\lambda}}  {\sqrt{2(N+1) \hat{\lambda} \lambda_{mn}}}\right)    
+ 
e^{2N}\left(
 \mathrm{erf} \left( \frac{N\lambda_{mn}-(N+1)\hat{\lambda}} {\sqrt{2(N+1) \lambda_{mn} \hat{\lambda}}} \right) 
- \mathrm{erf} \left(\frac{N\lambda_{mx}+(N+1)\hat{\lambda}} {\sqrt{2(N+1) \lambda_{mx} \hat{\lambda}}}\right)
\right)
 ) \\
 \approx( \hat{\lambda} \ln(\frac{\lambda_{mx}}{\lambda_{mn}}))^{-1}=p(\lambda) $, and therefore,\\
\noindent
$h(\hat{\Lambda})\approx h(\Lambda)$.

Remarks: (i)  Mathematica  performs the exact integration when one temporarily substitutes $m$ for $(N+1)$ and specifies the appropriate assumptions. (ii) With $N=2500$, and a naive use of Mathematica, $p(\hat{\lambda})$ approximations of $ p(\lambda)$ appears exact when  running at the default precision of Mathematica. The $\erf$ terms in $p(\hat{\lambda})$ combine to a value of two, which is the exact value needed (since this value is multiplied by one-half).
 
The second approximation concerns the conditional differential entropies.  Recall from Results,\\ 
$p(\hat{\lambda}|\lambda)=\sqrt{N+1} (2 \pi \lambda \hat{\lambda})^{-1/2}
\exp(-\frac{\lambda N^2}{2(N+1)\hat{\lambda}} - \frac{\hat{\lambda}(N+1)}{2\lambda}+N)$. Thus in bits and according to Mathematica,
$-h(\hat{\Lambda}|\lambda)=(\ln(2)^{-1}(\frac{1}{2}\ln(2\pi e(N+1)^2/N)-\ln(\lambda)+\frac{\exp(2N)\sqrt{N}\mathit{BesselK}^{(1,0)}[-\frac {1}{2},2N]} {\sqrt{\pi}})$. Before valuing this entropy to yield the second approximation, there is a simplification that obviates the need for a third approximation. When the expectation $h(\hat{\Lambda}|\Lambda)=\int p(\lambda) h(\hat{\Lambda}|\Lambda=\lambda)d\lambda$ is taken, the term $E[\ln(\Lambda)]$ appears. However, this same term of opposite sign appears in the marginal differential entropy $h(\Lambda)$, so they combine to zero. Then with $\frac{\exp(2\cdot 2500)\sqrt{2500}\mathit{BesselK}^{(1,0)}[-\frac {1}{2},2\cdot 2500]} {\sqrt{\pi}} < 10^{-3}\approx 0$, we have the second approximation, and this approximation improves as $N$ increases.
The information gain in continuous time (per sec) then is 
$h(\hat{\Lambda})-h(\hat{\Lambda}|\Lambda)\approx \log_2(\ln (\frac {\hat{\lambda}_{mx}}{\hat{\lambda}_{mn}})) + \frac{1}{2}\log_2(2\pi e(N+1)^2/N) $. 

\newpage

\FloatBarrier
%\subsection*{Energy Audit Overview }

\begin{figure}

\textbf{Energy Audit Overview Including Partitioning }\par\medskip
 \includegraphics[scale=0.55]{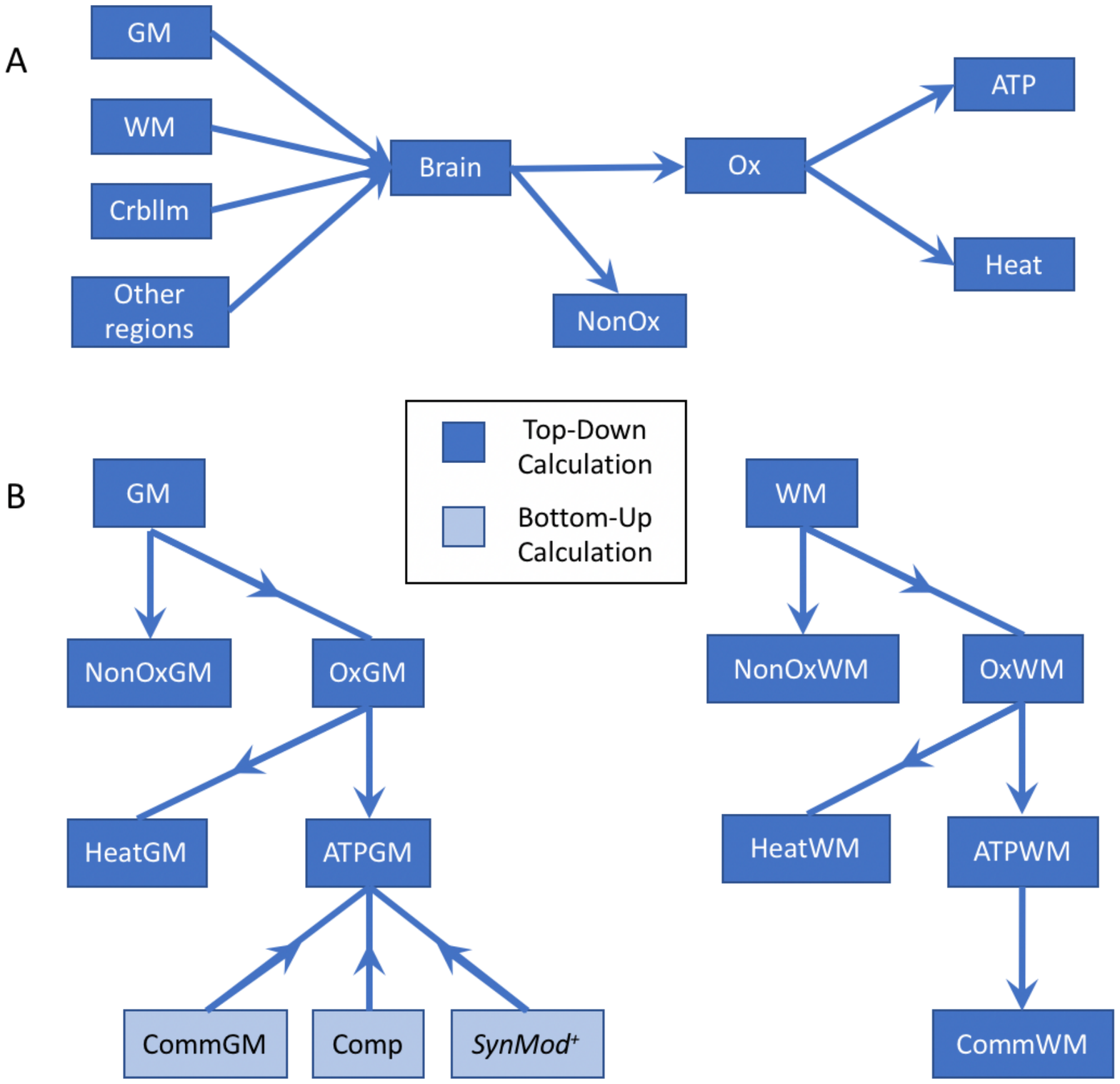}
\caption{The partitioning of the energy available from glucose for the top-down estimates. Note: Arrows indicate our partitioning process, not flow of energy/glucose. A: Proceeding from left to right, total brain glucose-uptake is calculated by summing the regional uptakes. From there, the total glucose is subdivided by metabolic fate. B: We begin with the regional cortical gray matter glucose uptake and end with specific ATP consumption. The first level of partitioning combines regional rates of [$^{11}$C]glucose uptake with regional brain weights; at the second level, glucose is partitioned based on metabolic fate (OxGM, the glucose that goes into cellular respiration, and NonOxGM, the glucose that is not oxidized); at the third level, energy from oxidized glucose is partitioned between ATP production and heat generation, and the majority of glucose-energy goes to the latter. At the fourth level, ATP energy is partitioned between communication, computation, and $SynMod^+$. GM-gray matter; WM-white matter; Crbllm-cerebellum; Other regions, see Table S1; NonOx-nonoxidized glucose; Ox-oxidized glucose; Comm-communication; Comp-computation; Other-includes synaptic modification, growth/retraction, maintenance, etc.}
\label{partitioning}
\end{figure}

\FloatBarrier
\newpage
%\subsection*{Glucose Partitioning:  by region and by metabolic fate}

\begin{table}[htbp]%
\centering
\begin{threeparttable}
\caption{Glucose Partitioning}
\label{glupartitioning} \centering %
\begin{tabular}{llll}
\toprule
& Mass (g) \tnote{$\diamond$} & Glucose uptake \tnote{$\diamond\diamond$} & Glucose uptake \tnote{$\star$}\\
& & $\mu mol/100g/min $ & $\mu mol/$region$/$sec\\
 \toprule %
whole brain & 1495 & – &  6.05\tnote{$\star\star$} \\

 \,forebrain cortex\\
\; cortical gray  &  \;633  &  28.6  &3.02\\
\, cortical white &  \;590  &  18.4   & 1.81  \\
\;cerebellum  &  \;154  &  24.7  &0.63  \\ 
\;other regions\,\tnote{$\dagger $} & \;118 & 30\tnote{$\dagger\dagger$} &0.59\\
\;ventricular\tnote{$\ddagger$}  &   5-15 & 0\\

\bottomrule 
\end{tabular}
\begin{tablenotes} \footnotesize
\item[$\diamond$] Regional masses from \cite{azevedo2009equal}. 
\item[$\diamond\diamond$] values from \cite{graham2002fdg}
\item[$\star$]Individual regional values are calculated from first two columns. 
\item[$\star\star$] This is a sum of the regional values. See text for details.
\item[$\dagger$]Includes basal ganglia, thalamus, hypothalamus, etc.
\item[$\dagger\dagger$] Uses avg. glucose uptake value of striatum and thalamus of \cite{graham2002fdg}
\item[$\ddagger$]This range is based on the possible remaining mass using values from \cite{azevedo2009equal}
\end{tablenotes}
\end{threeparttable}
\end{table}

%\FloatBarrier

\begin{table}[!ht]%
\centering
\begin{threeparttable}
\caption{Glucose Energy Partitioning to ATP-watts\tnote{$\diamond$}}
\label{mainwattsnathandmitchell} \centering %
\begin{tabular}{lllll}
\toprule
Top-Down Calculations & Watts\tnote{$\diamond$} & Non-oxidized\tnote{$\diamond\diamond$}  & ATP- & ATP-\\
&(complete  & (equivalent & watts$^{\star}$ &  watts$^{\star\star}$ \\
 & oxidation) & watts)  & & \\\toprule %
whole brain (1495 g)& 17.0 & 1.86 & 6.19 & 5.23\\
\midrule
\;\;cerebellum (154 g)  & 1.77 & 0.19  & 0.65  & 0.55 \\ 
\;\;other regions\,  (118 g)\tnote{$\dagger $} & 1.65 & 0.18 & 0.60  & 0.51\\
\midrule
\;\;forebrain cortex (1223 g):\\
\;\;\;\;white (590 g)  & 5.07 & 0.56 &1.85 & 1.56\\
\;\;\;\;gray \, (633 g) & 8.45 & 0.93 &3.09  & 2.61\\
\midrule
Bottom-up calculations\\
\midrule

\;\;\;\;gray: 1.0 Hz\\
\quad\; communication & &  & 1.68 & 1.68\\
\quad\; computation  & & & 0.10 &  0.10\\
\quad\; $\textit{SynMod}^+$ & & &  1.31\tnote{$\ddagger$}  &  0.84\tnote{$\ddagger$}  \\

\;\;\;\;gray: 1.6Hz\\
\quad\; communication & &  & 2.75 & 2.75\tnote{$\dagger\dagger$}\\
\quad\; computation  & & & 0.17 &  0.17\tnote{$\dagger\dagger$}\\
\quad\; other ATP demands & & & 0.17\tnote{$\ddagger$}  & 0  \\

\midrule
\bottomrule 
\end{tabular}
\begin{tablenotes} \footnotesize
\item[$\diamond$]Watts based on glucose-uptake values from \cite{graham2002fdg} and $2.8$ $MJ/mol\;glucose$ \cite{nelson2008lehninger} and $36$  $kJ/molATP$ \cite{nath2016thermodynamic}\cite{nath2009energy}; regional masses from \cite{azevedo2009equal}. 
\item[$\diamond\diamond$] Also assuming complete oxidation of glucose. See partitioning of glucose in earlier sections
\item[$\star$]Using Nath's torsional mechanism \cite{nath2016thermodynamic}, \cite{nath2010beyond}  which incorporates mitochondrial leak 
\item[$\star\star$]Using chemiosmotic theory \cite{mitchell1966chemiosmotic} which is then downgraded by standard mitochondrial leak value: 20\%  \cite{rolfe1997cellular}
\item[$\dagger$]Including basal ganglia, thalamus, brainstem, etc. The missing mass is ventricular. See Table \ref{glupartitioning} and the accompanying footnotes for more information.
\item[$\dagger\dagger$]Indicates bottom-up values exceed available energy if the top-down calculations are accepted.
\item[$\ddagger$] Assuming that $SynMod^+$ consumes the remaining gray matter ATP-watts ($SynMod^+$ = gray matter ATP-watts - communication - computation).
\end{tablenotes}

\end{threeparttable}
\end{table}

\FloatBarrier

%\FloatBarrier
\quad
\newpage
%\subsection*{Comments on data selection from the  literature and inherent imprecisions}
%\subsection*{Bottom-up values and intermediate calculations}
\begin{table}[htbp]%
\centering
\begin{threeparttable}
 \caption{Bottom-Up Computation and Communication$^\star$}
\label{results}\centering 
%\begin{tabular}{l}
%\hspace{33pt}(watts/GM or joules/neuron(NRN)/AP)
%\end{tabular}
\begin{tabular}{lll}
   \toprule
%\multicolumn{3}{c}{Gray Matter(GM) Computational cost}\\ 
%\multicolumn{3}{c}{}\\
%\toprule
\multicolumn{3}{c}{Gray Matter Computational Costs$^\diamond$}  \\
  \midrule 
AMPAR  & NMDAR$^{\dagger}$   &  AMPAR + NMDAR  \\
0.069 W & $0.5\cdot 0.069$  $\approx 0.035$ W & 0.104 W\\
\toprule
\multicolumn{3}{c}{Gray Matter Communication 
Costs }  \\
 \midrule 
 Resting Potential  & Action Potentials  &  Presynaptic Transmission \\
 1.09 W & 0.47 W & 0.12 W \\
  
 $ 7.3\cdot 10^{-11}$ J/NRN/AP & $ 3.1 \cdot 10^{-11}$ J/NRN/AP & $ 7.3\cdot 10^{-12}$ J/NRN/AP \\
 
 \toprule
 \multicolumn{3}{c}{Gray Matter Communication + Computation}\\
 \midrule
GM Comm Total & Comp Total & Comm+Comp  \\
 1.68 W &  0.10 W & 1.78 W\\
 $1.11\cdot 10^{-10}$  J/NRN/AP & $6.67\cdot 10^{-12}$ J/NRN/AP 
 &$1.18 \cdot 10^{-10}$ W/NRN \\
 \toprule
  \multicolumn{3}{c}{Other Totals (cortex = WM + GM)}\\
 
 \toprule 
 WM\tnote{$\star\star$}\;\; + GM Comm &Comp + \textit{SynMod}$^+$\tnote{$\diamond\diamond$} & Total GM + WM    \\
  3.53 W/cortex  &  1.41 W   & 4.94 W/cortex   \\
 $2.35\cdot10^{-10}$ J/NRN/AP & $2.13\cdot10^{-10}$ J/NRN/AP& $3.29\cdot10^{-10}$ J/NRN/AP\\

\bottomrule
\end{tabular}

\begin{tablenotes} \footnotesize
\item[$\star$]For more details on these calculations, see Methods

\item[$\diamond$] $1.0$ Hz with $75\%$ failure rate implies 0.4 SA/synapse/sec
\item[$\dagger$] Assumes NMDA-receptor activation contributes half-again the cost of AMPA-receptor activations
\item[$\star\star$] White matter communications also includes white matter \textit{SynMod}$^+$
\item[$\diamond\diamond$] Gray matter \textit{SynMod}$^+$ includes synaptic modification costs, e.g., metabotropic transmitter effects, 
axonal growth/retraction, receptor modification/removal as well as the transport needed for such modifications

\end{tablenotes}
 
\end{threeparttable}
\end{table}

\FloatBarrier
\quad
\clearpage

%\subsection*{Computation Costs}
%\FloatBarrier

\begin{table}[htbp]%
\centering
\begin{threeparttable}
\caption{Computational costs arising from ionotropic glutamate synaptic activations}
\label{comp}\centering %
\begin{tabular}{lllll}
   \toprule
\multicolumn{5}{c}{ Voltages }  \\
   \midrule
 $V_{rev}$ & $V_{m,ave}$ & $V_{Na,Nern.}$& $V_{K,Nern}$ &\\ 
 $-7$ mV & $-55$ mV &$+55$ mV &$-90$ mV \\
  \midrule
\multicolumn{5}{c}{For a single synaspe, Average (Ave) AMPA-Receptor (AR) per  synaptic activation (SA)\tnote{$\star$} }  \\
   \midrule
$G_{ave}$ & $(V_{Na^+}-V_{ave,m})\cdot G_{Na}/$SA &  $Na^+$ amps & Na$^+$ coulombs/SA  \\ 
$200$ pS/SA   & \quad $110$ mV $\cdot114.5$ pS/SA  &   12.5 pA/SA    &  15.1 fC/SA\\
 \toprule
 \multicolumn{5}{c}{AMPAR failure rate-adjusted costs\tnote{$\diamond$} (All GM synapses$=1.5\cdot 10^{14}$)  }  \\
 \midrule 
 $Na^+$flux &  $Na^+$flux  & ATP mol/sec & ATP watts \\
 0.057 C/sec  & $5.87\cdot 10^{-6}$ mol/sec & $1.96\cdot 10^{-6}$ mol/sec & 0.069 W \\
 \toprule 
\multicolumn{5}{c}{AMPAR + NMDAR Computational cost}  \\
  \midrule 
per cerebral cortex 
& & per neuron per spike  \\
$1.5^{\dagger}\cdot 0.069$ W $\approx 0.10$ W
& & $7.0 \cdot 10^{-12}$ J \\
\bottomrule
\end{tabular}
\begin{tablenotes}
\footnotesize
\item{$\hspace{-0.22cm} ^\star$} SA duration  $1.2$ msec 
\item[$\diamond$] $1.0$ Hz with $75\%$ failure rate
\item[$\dagger$] Assumes NMDA-receptor activation contributes half-again the cost of AMPA-receptor activations
\end{tablenotes}
\end{threeparttable}
\end{table}

\FloatBarrier
\clearpage

%\subsection*{Communication Costs}

\begin{table}[!ht]%
\centering
\begin{threeparttable}
\caption{ Gray Matter Communication\tnote{$\diamond$}}
\label{comm1.6}\centering %
\begin{tabular}{llll}

 \toprule
\multicolumn{4}{c}{Parameters. I (all conductances are at rest)} \\
 \midrule 
$V_m^{rest}$ & $V_{Na,Nern}$ & $V_{K,Nern}$ &$G_{Na}:G_K$ \\
$-66\,mV$ & $+55\,mV$ & $-90\,mV$ & $121:24$   \\
 \toprule
\multicolumn{4}{c}{Parameters. II} \\
 \midrule 
 Capacitivity & Resistivity & Conductivity & Axon/Bouton Area \\
$0.96\,\mu F/cm^2$  &  $50,000\, \si{\ohm} cm^2$  & $2.00\cdot 10^{-5}\,S/cm^2$ &$\approx2:1$\\
 \toprule
\multicolumn{4}{c}{Total across all axons or boutons} \\
 \midrule 
Area Axon $+$ Bouton  & $G_{axon+btn}^{rest}$ & $G_{Na}^{rest}$ &$C_{ax};\quad C_{btn}$\\
$21.8\cdot 10^6 \,cm^2$ & $436 \,S$ & $72\, S$ &
$14.6\,F;\;6.34\,F$\\
 \toprule
\multicolumn{4}{c}{$Na^+$ rest-flux and ATP to remove}  \\
\midrule
 $Na^+$ flux & $Na^+$ flux   & ATPs used & \\

 $8.73\,C/s$ & $9.05\cdot 10^{-5}\, mol/s$ & $3.02\cdot 10^{-5}\,mol/s$ \\
 \toprule
\multicolumn{4}{l}{Cost of axon $+$ bouton rest potentials ($36000\,J/molATP$ )}
1.09 W\\
 \midrule
 \toprule
\multicolumn{4}{c}{110mV ($V_{AP}$) Axon Action Potential (AP); $20\; mV$ ($\Delta V_{Bt}$) Bouton depolarization}  \\
 \midrule
 Axon charging/sec\tnote{$\diamond$} & Overlap-scaled\tnote{$\star$} & Bouton charging/sec\tnote{$\diamond\diamond$} &$Na^+ mol/sec $ \\
   $ 1.61$ amps& $3.66$ amps & $ 0.13$ amps & 
   $3.92 \cdot 10^{-5}$ mol/sec\\

 \toprule
\multicolumn{4}{l}{Cost of action potentials ($1.31 \cdot 10^{-5}\; ATP \, mol/sec$)}  
0.47 W\\
 \midrule

 \toprule
\multicolumn{4}{c}{ ATP  per vesicle released (VR); ATP per $Ca^{2+}$  spike}\\
 \midrule
$ATP/vesicle$ &$Ca^{+2} \, removal$  & \\
$5.7\cdot 10^{3}$ ATPs & $1.2\cdot 10^{4}$ATPs  &   \\
$(9.5\cdot 10^{-21}$ mols) &$(1.99\cdot 10^{-20}$ mols) &  \\
 \toprule
\multicolumn{4}{c}{Presynaptic AP costs \tnote{$\star\star$}}\\
 \midrule
VR Events/s &Ca events/s & VR ATP use & $Ca^{2+}$ ATP use \\
$3.75 \cdot 10^{13}$/s & $1.5\cdot 10^{14}/s$ &$3.57\cdot 10^{-7}$mol/s & $2.99\cdot 10^{-6}$ mol/s\\

 \midrule

 \toprule
\multicolumn{4}{l}{AP Generated Presynaptic Cost} 0.12
W\\
\midrule  

 \toprule
\multicolumn{4}{l}{Total} 1.68  W\\
 \midrule

\bottomrule
\end{tabular}

 \begin{tablenotes} \footnotesize
 \item[$\diamond$] 1.0 Hz firing rate 
 \item[$\star$] Multiplier from \cite{hallermann2012state} 
 \item[$\star\star$] 75\% failure rate of TR but no failure of Ca entry
\end{tablenotes}
\end{threeparttable}
\end{table}

\FloatBarrier
\quad 
\clearpage

\begin{table}[htbp]%
\centering
\begin{threeparttable}
\caption{GM Communication Volume Fractions and Cylindrical Approximations\tnote{$\diamond$}}
\label{surfacearea}\centering %
\begin{tabular}{llllllll}
\toprule
 & Vol. Frac & Volume \tnote{$\diamond\diamond$}& Diameter& Height\tnote{$\star$} & Length\tnote{$\star\star$} &$ Area_{pm}$\tnote{$\dagger$}   & C$_m$\tnote{$\dagger\dagger$} \\
 & & & & per bouton & (total) \\\toprule
Boutons&22$\%$ &139 cm$^3$& 1.1$\;\mu $m& 1.0$\;\mu $m&  - &6.61$\cdot 10^6$ cm$^2 $ & $6.34$ F \\
Axons & $30\%$ & 190 cm$^3$  &0.50 $\mu $m & - & 9.66$\cdot 10^{10}$ cm   &  $15.2\cdot 10^6$ cm$^2$ &$14.6$ F\\
Total & 52$\%$ & 329 cm$^3$ & - & - & - & $21.8\cdot 10^6$ cm$^2$ & $20.9$ F\\

\bottomrule
\end{tabular}

 \begin{tablenotes} \footnotesize
 \item[$\diamond$] Assuming $1.5\cdot 10^{14}$ synapses per cortex  \item[$\diamond\diamond$]Cortical gray matter volume \item[$\star$]Height is for a single cylindrical bouton 
 \item[$\star\star$]Length is total length of all small axons
 \item[$\dagger$]Area of plasma membrane (pm) \item[$\dagger\dagger$]Membrane capacitance using $0.96\;\mu$F/cm$^2$ based on $0.88\;\mu$F/cm$^2$ of the lipid membrane plus activation of two-thirds of the $0.12\;\mu$F/cm$^2$ of Na$^+$channel gating-charge. 

\end{tablenotes}
\end{threeparttable}
\end{table}

\FloatBarrier

\FloatBarrier

\begin{table}[htbp]%
\centering
\begin{threeparttable}
\caption{Other axon parameter sets consistent with a 30\% volume-fraction}
\label{SAandcomm}\centering %
\begin{tabular}{llllll}
\toprule
Axon Diameter ($\mu m$) & 0.28 & 0.4 & 0.5 & 0.6 \\
\midrule
 Axon Area $(cm^2)$& $27.1\cdot 10^6$ & $18.9\cdot 10^6$& $15.2\cdot 10^6$ & $12.7\cdot 10^6$\\ 

 Total Area\tnote{$\diamond$}\; $(cm^2)$ & $33.7\cdot 10^6$ & $25.6\cdot 10^6$ & $21.8\cdot 10^6$  & $19.3\cdot 10^6$ \\
 Axon Length\tnote{$\diamond\diamond$} \; $(cm)$ & $3.08\cdot 10^{11}$ & $1.51\cdot 10^{11}$ & $9.67\cdot 10^{10}$ &  $6.72\cdot 10^{10}$ \\
 RP cost (W) & 1.67 & 1.28 & 1.09 & 0.96\\
 AP cost (W) & 0.83 & 0.58 & 0.47  & 0.39\\

\bottomrule
\end{tabular}

 \begin{tablenotes} \footnotesize
 \item[$\diamond$] Total area is the sum of axonal surface area and bouton surface area as in Table \ref{surfacearea}, 22\% volume-fraction.
 \item[$\diamond\diamond$] The axon length is the length of all axons extended by 1.0 $\mu$m assumed  bouton height and number.
\end{tablenotes}
\end{threeparttable}
\end{table}

\FloatBarrier

\FloatBarrier

\begin{table}[htbp]%
\centering
\begin{threeparttable}
\caption{Effect of Varying Bouton Dimensions}
\label{boutons}\centering %
\begin{tabular}{lllll}
\toprule
  Diameter& Height &  $Area_{pm}$  & Vol. Frac & C$_m$ \\\toprule
0.9$\mu m$& 1.0$\mu m$ & $5.19\cdot 10^6 cm^2 $ & 15$\%$ &  $ 4.99F$ \\
0.9$\mu m$ & 1.2$\mu m$ &  $ 6.04\cdot 10^6 cm^2$ & 18$\%$  &$ 5.80F$\\
 $1.1\mu m$ & $1.0\mu m$ &  $ 6.61\cdot 10^6 cm^2$ & 22$\%$ & $ 6.34F$\\

\bottomrule
\end{tabular}

\end{threeparttable}
\end{table}

 \newpage

\end{document}